\documentclass[prb,twocolumn,superscriptaddress,longbibliography]{revtex4-1}
\RequirePackage[mathlines]{lineno}
\usepackage{amssymb}
\usepackage{amsmath}
\usepackage{mathrsfs}
\usepackage{graphicx}
\usepackage[usenames,dvipsnames]{xcolor}
\definecolor{goodgreen}{rgb}{0.1,0.5,0}
\definecolor{goodred}{rgb}{0.7,0,0}
\usepackage[colorlinks,urlcolor=goodgreen,citecolor=blue,linkcolor=goodred]{hyperref}

\makeatletter
\newsavebox{\@brx}
\newcommand{\llangle}[1][]{\savebox{\@brx}{\(\m@th{#1\langle}\)}%
  \mathopen{\copy\@brx\kern-0.5\wd\@brx\usebox{\@brx}}}
\newcommand{\rrangle}[1][]{\savebox{\@brx}{\(\m@th{#1\rangle}\)}%
  \mathclose{\copy\@brx\kern-0.5\wd\@brx\usebox{\@brx}}}
\makeatother

\usepackage[normalem]{ulem}

\newcommand{\dcom}[1]{\textcolor{NavyBlue}{\textsf{\textbf{#1}}}}

\newcommand{\vect}[1]{\boldsymbol{#1}}

\setpagewiselinenumbers
\modulolinenumbers[5]

\begin{document}

\preprint{\dcom{Preprint not for distribution CONFIDENTIAL, Version of \today}}
\title{Chiral correlation of drag currents inducing optical activity of twisted bilayer graphene}

\author{S. Ta Ho}
\affiliation{National University of Civil Engineering (NUCE), 55 Giai Phong road, Hanoi 10000, Vietnam}
\author{V. Nam Do}
\email{nam.dovan@phenikaa-uni.edu.vn}    
\affiliation{Department of Basic Science, Phenikaa Institute for Advanced Study (PIAS), A1 Building, Phenikaa University, Hanoi 10000, Vietnam}

\begin{abstract}
The mechanisms of optical activity and quantum transport of twisted bilayer graphene are studied. The formation of unique electron states in the bilayer systems is studied using an effective continuum model. Such states are shown to support the correlation of transverse motions of electrons in two graphene layers. Because of the chiral structure of the atomic lattices, the contribution of such drag correlations is incompletely cancelled, thus resulting in a drag term of the optical conductivity tensor. We show that the drag term of the conductivity is the manifestation of the spatial dispersion. We show how to analyze and calculate the components of the conductivity tensors that governs the optical activity of the systems. The DC conductivity of the twisted bilayer graphene system is also calculated. It shows the existence of a quantum conductivity value $\propto e^2/h$ at the intrinsic Fermi energy.
\end{abstract}
\maketitle

\section{Introduction}
In recent years, research on engineering two-dimensional (2D) materials has been developed intensively and extensively. Achievements shows great potentials of using multiple layer materials to applications in electronics, optoelectronics,  spintronics and valleytronics.\cite{Sierra_2021,Liao_2019,Sangwan_2018,bao_2019} Using multiple layer materials with the van der Waals (vdW) interlayer coupling brings about many solutions to tune the electronic structure, e.g., changing the interlayer distance, changing the lattice alignment by twisting and/or sliding constitution layers and applying external fields. Physically, such solutions of material engineering typically cause the breaking of spatial and temporal symmetries that govern the dynamics of electrons confined inside material layers. For example, a magnetic field breaks the time reversal symmetry. It results in many intriguing phenomena such as the quantum Hall and magneto-optical effects, including the Kerr and  Faraday rotation of the polarization plane of linearly polarized light as well as the circular dichroism.\cite{Tse_2011,Crassee_2011,Nandkishore_2011,Shimano_2013} Such striking optical phenomena have been exploited to develop devices in many areas of application.\cite{Izake_2007,Liu_2005,Solomon_2019} However, using external fields does not allow designing small-scale devices. Accordingly, searching materials with intrinsic properties and solutions to break spatial symmetries of an electronic system to match technical requirements is paramount.

Twisted-bilayer graphene (TBG) is a typical 2D vdW material that has been attracting an intensive study from research community.\cite{Andrei_2020,Nimbalkar_2020,Hill_2021} The TBG system is created by stacking two graphene layers on each other such that the alignment between two hexagonal lattices is characterized by a rotation, or twist, angle. Experimental researches showed the ability to tune the twist angle mechanically.\cite{Palau_2018,Cai_2021} For the electron system inside the TBG lattices the time reversal symmetry is not broken. The TBG lattices only process rotational symmetries of the group $D_{3}$ or $D_6$.\cite{Zou_2019,Do_2020} As a consequence, the response of the TBG systems to an electromagnetic field is characterized by a diagonal optical  conductivity tensor whose elements are $\sigma_{xx}(\omega)=\sigma_{yy}(\omega)\neq 0$ and  $\sigma_{xy}(\omega)=0$. However, the TBG systems were pointed out to support the circular dichroism phenomenon.\cite{Kim_2016} It was generally supposed to lie in the fact that the TBG atomic lattices is chiral, i.e., the left-handed and right-handed twist configurations are not the mirror images of each other. The optical activity of crystalline solids can be generally explained as a magneto-electric effect.\cite{Condon_1937,Kim_2016,Nandkishore_2011,Stauber-2018} However, microscopic mechanisms for the optical activity of diverse materials are not unique. It is thus desirable to clarify a particular mechanism for novel multiple layer vdW materials in general and for twisted bilayer graphene in particular.

Microscopically, Morell employed an effective continuum model wherein the dynamics of electrons is described as the coupling of Dirac fermions in each graphene layer.\cite{Morell_2017} The Dirac fermions are characterized by a helicity. Twisting two graphene layer makes the non-colinear of the two helicities in two layers. Combined with a symmetry analysis, they discussed that the origin of the observed CD spectrum is due to the misalignment of the helicities. Independently, Do et. al. performed a numerical study on the basis of a tight-binding model to simulate the evolution of electron states in the whole TBG lattices.\cite{Do_2020} Initial states are created as excitations on different lattice nodes of graphene layers. They showed a spreading of the electron wave functions not only in the atomic lattice, but also the interchange between the two layers. It was shown that special excited states can similarly occur to the picture of neutrino oscillations.\cite{Xiao_2013,Do-2019} Especially, due to the misalignment of two hexagonal lattices, the wave spreading in one graphene layer results in the spreading in the other layer with a twist of the wavefront shape. Effectively, a motion of electron in one layer induces a transverse motion in the other layer, forming a picture similar to that of the classical Hall effect. The simulation results of the evolution of Gaussian wave packets show the trajectory of the wave centroid in curly curves, deflecting from the initial direction of the wave packets. Numerical calculation shows a non-zero Hall conductivity, demonstrating the existence of correlation between the longitudinal and transverse components of the current densities in two graphene layers. This result seems to violate the symmetry properties of thermodynamics like the Onsager reciprocal relations.\cite{Groot_1984} However, it is totally not because it is the result only from a particular state excited from one layer. An opposite behavior will occur if one excites an initial state on the other graphene layer. Eventually, taking average over an ensemble of possible initial states will exactly result in a total zero transverse conductivity $\sigma_{xy}(\omega)$. That simulation is consistent with optical experimental setups wherein incident photons know which graphene layer they hit first and which one they hit later. The non-zero Hall conductivity supported by individual electron states in the TBG lattices therefore should be understood as the drag Hall conductivity. Accordingly, in order to correctly study the optical response of multiple layer vdW systems the spatial dispersion along the direction perpendicular to material layers must be appropriately taken into account.\cite{Agranovich_2013,Poshakinskiy_2018}

In this paper we present a theory for analyzing the response of multiple layer vdW systems to external electromagnetic fields by explicitly taking into the spatial dispersion. This theory naturally leads to the decomposition of the conductivity tensor into two parts. The first part, called the local one, defines the local relation of the current density and the electric field in each layer. The second part, however, defines the linear relation of the current density in one layer to the value of the electric field in other layers. The second part therefore encodes the non-local effects and is called the drag one. We will show how the elements of these two conductivity parts are calculated using the Kubo formula. We then apply this theory to the TBG systems and show how the local and drag conductivities play their role in governing the chiral response of the material systems. For this, we have to solve the problem of propagation of an electromagnetic wave through a TBG sheet without ignoring its thickness though small. It means that we explicitly consider each graphene layer as an interface of a material slap separating two dielectric mediums. These interfaces, however, support two current densities which are mutually influenced to each other, different from ordinary dielectric slaps. The mutual relation is governed by the interlayer coupling. The hybridization of electronic states in two graphene layers results in unique states in layer-coupled systems. Interestingly, we will show that these hybridized states also support a quantum value of the DC conductivity at the intrinsic Fermi energy.

This article has three main sections. Aside from the introduction, in the next section, Sec. II, we present the theory for decomposing the total conductivity tensor into two parts and show how they are calculated via the Kubo formula. Subsequently, we present a solution to an effective model for electrons in the TBG lattices. In Sec. III, we present numerical results for the electronic band structure of several TBG configurations, results for the analysis of the optical conductivity tensor elements, and the spectra of circular dichroism with respect to photon frequency. Finally, the conclusion is given in section IV.

\section{Theory and calculation methods}
\subsection{AC and DC conductivities}
The optical properties of bilayer graphene systems are noticeable.\cite{Moon-2013,Stauber-2018,Do_2018,Catarina_2019} In particular, when the layer resolution should be taken into account, effects involved in the spatial dispersion would emerge. To describe such effects, we decompose the current responding to the action of an external electric field of a monochromatic light beam incident perpendicular to the material plane as follows:
\begin{align}\label{Eq1}
    J_\alpha(\vect{x},z, \omega) &= \int dz^\prime\sigma_{\alpha\beta}(\vect{x},z;\vect{x},z^\prime,\omega)E_\beta(\vect{x},z^\prime,\omega).
\end{align}
On the assumption of spatial homogeneity in the material plane, we can write $\sigma_{\alpha\beta}(\vect{x},z;\vect{x},z^\prime, \omega) = \sigma_{\alpha\beta}(z,z^\prime,\omega)$, and:
\begin{align}\label{Eq2}
    \sigma_{\alpha\beta}(z,z^\prime,\omega)  =&\sum_{\ell=1}^2\sigma_{\alpha\beta}(z,z^\prime,\omega)\delta(z^\prime-z_\ell).
\end{align}
Eq. (\ref{Eq1}) therefore becomes:
\begin{align}\label{Eq3}
    J_\alpha(\vect{x},z,\omega) = \sum_{\ell=1}^2\sigma_{\alpha\beta}(z,z_\ell,\omega)E_\beta(\vect{x},z_\ell,\omega).
\end{align}
More specifically, we resolve the current density induced in each graphene layer as follows:
\begin{equation}\label{Eq4}
    J_\alpha^{(\ell)}(\vect{x},\omega) = \sigma_{\alpha\beta}^{(\ell)}(\omega)E_{\beta}^{(\ell)}(\vect{x},\omega)+\sigma_{\alpha\beta}^{(drag)}(\omega)E_\beta^{(\ell^\prime)}(\vect{x},\omega),
\end{equation}
where the involved quantities are defined by: $J_\alpha^{(\ell)}(\vect{x},\omega) = J_\alpha(\vect{x},z_\ell,\omega); E_\beta^{(\ell)}(\vect{x},\omega) = E_\beta(\vect{x},z_\ell,\omega)$ and $\sigma_{\alpha\beta}^{(\ell)}(\omega) = \sigma_{\alpha\beta}(z_\ell,z_\ell,\omega)$, and  $\sigma_{\alpha\beta}^{(drag)}(\omega) = \sigma_{\alpha\beta}(z_\ell,z_{\ell^\prime},\omega)$ where $\ell\neq\ell^\prime$. Eqs. (\ref{Eq4}) can be cast in the matrix form:
\begin{align}\label{Eq5}
    \left(\begin{array}{c}
    \vect{J}^{(1)}\\ \vect{J}^{(2)}
    \end{array}\right)_{\omega} = \left(\begin{array}{cc}
    \vect{\sigma}^{(1)} & \vect{\sigma}^{(drag)}\\
    \vect{\sigma}^{(drag)\dagger} &\vect{\sigma}^{(2)}
    \end{array}\right)_\omega\left(\begin{array}{c}
    \vect{E}^{(1)}\\ \vect{E}^{(2)}
    \end{array}\right)_{\omega},
\end{align}
where $\vect{\sigma}^{(\ell)}(\omega) = \sigma^{(\ell)}(\omega)\vect{\tau}_0$ is called the local part, and  $\vect{\sigma}^{(drag)}(\omega) = \sigma^{(drag)}(\omega)\vect{\tau}_0-i\sigma_{xy}^{(drag)}(\omega)\vect{\tau}_y = \sigma^{drag}(\omega)-\sigma_{xy}^{drag}\hat{\vect{z}}\times$ is called the drag part. Here $\vect{\tau}_0$ is the $2\times 2$ identity matrix and $\vect{\tau}_2$ is the conventional second Pauli matrix. The compact form (\ref{Eq5}) is identical to results reported in Refs. \onlinecite{Stauber-2018,Ochoa_2020}. Importantly, it guarantees the time-reversal symmetry, the rotation symmetry and the layer interchange symmetry. 

The conductivities $\sigma^{(\ell)}(\omega), \sigma^{(drag)}(\omega)$ and $\sigma^{(drag)}_{xy}(\omega)$ can be determined from the Kubo formula. There are a number of versions of the Kubo formula for the electrical conductivity suitable for implementing it in different situations. However, the most important ingredient we must specify is the velocity operators $\hat{v}_\alpha$. In general, these operators are determined from the position operator $\hat{x}_\alpha$ and the Hamiltonian $\hat{H}$ via the Heisenberg equation:
\begin{equation}\label{Eq6}
    \hat{v}_\alpha = \frac{1}{i\hbar}[\hat{x}_\alpha,\hat{H}] \rightarrow v_\alpha(\vect{k})= \frac{1}{\hbar}\frac{\partial}{\partial k_\alpha}H(\vect{k}).
\end{equation}
According to the linear response theory, we see that the conductivity has the paramagnetic part only. Using the eigen-vectors of single-particle Hamiltonian (without the presence of external vector potential $\vect{A}(\vect{x},t)$) as the representation basis, $\hat{H}|n\rangle = E_n|n\rangle$, the elements of the optical conductivity tensor are given by this formula:~\cite{Allen_2006}
\begin{align}\label{Eq7}
\sigma_{\alpha\beta}^{(c)}(\omega) &= \frac{i\hbar  e^2}{S}\sum_{m,n}\frac{f_n-f_m}{E_m-E_n}\frac{O^{(c)mn}_{\alpha\beta}}{\hbar(\omega+i\eta)-(E_m-E_n)},
\end{align}
where $S$ is the area of material sample; $c = \{\ell,drag\}$; $f_n = f(E-\mu,k_BT)$ is the occupation weight of the energy level $E_n$ determined by the Fermi-Dirac function $f$ with $\mu,k_BT$ the chemical potential and thermal energy; $\eta$ is a positive infinitesimal number, and $O_{\alpha\beta;mn}^{(c)}$ denotes the product of velocity matrix elements:
\begin{subequations}
\begin{align}
    O_{\alpha\beta}^{(\ell)mn} &= \langle m|\hat{v}_\alpha^{(\ell)}|n\rangle\langle n|\hat{v}_\beta^{(\ell)}|m\rangle,\label{Eq8a} \\
    O_{\alpha\beta}^{(drag)mn} &= \langle m|\hat{v}_\alpha^{(1)}|n\rangle\langle n|\hat{v}_\beta^{(2)}|m\rangle\label{Eq8b}.
\end{align}
\end{subequations}

To the DC conductivity, it might be optimistic to think that the DC conductivity can be simply obtained from the expression of the optical conductivity in the limit of zero frequency. However, it is not, at least in the numerical calculation. This is because for some finite frequency $\omega$, there is always a finite length scale governing the behavior of electron. This length scale is $L_\omega = 2\pi v_F/\omega$. Meanwhile, there is no such length scale for the DC transport. Furthermore, the static transport has the diffusion nature due to the similarity of the Schr\"odinger equation and the diffusion equation. The DC conductivity can be obtained from the linear response theory. The vector potential is chosen in the form $\vect{A}(t) = (-Et,0,0)$, where $E$ is the intensity of a static electric field. The DC conductivity is calculated using the Kubo-Greenwood formula:\cite{Kubo_1957,Greenwood_1958}
\begin{equation}\label{Eq9}
    \sigma_{\alpha\alpha}(\mu,kT) = -\int_{-\infty}^{+\infty}dE \frac{\partial f(E-\mu,kT)}{\partial E}\sigma_{\alpha\alpha}(E),
\end{equation}
where
\begin{equation}\label{Eq10}
    \sigma_{\alpha\alpha}(E) = \frac{2\pi e^2\hbar}{S}\sum_{m,n}|\langle m|\hat{v}_\alpha|n\rangle|^2 \delta(E-E_n)\delta(E-E_m).
\end{equation}
This quantity is seen as the DC conductivity at zero temperature. When performing the above formula, we use the Gaussian function to approximate the $\delta$-Dirac function:
\begin{equation}\label{Eq11}
    \delta(E-E_n) \approx \frac{1}{\eta\sqrt{\pi}}\exp\left(-\frac{(E-E_n)^2}{\eta^2}\right),
\end{equation}
where $\eta > 0$ is a tiny number that is appropriately chosen to smear the energy levels for the numerical calculation.
\subsection{Effective continuum model}
Developing effective continuum models for low-energy states of electrons in TBG systems has been proceeded since 2007 by Lopes et al.\cite{Santos_2007} However, the model driven by Bistritzer and MacDonald in 2011 was well known and commonly used. Here, we present our solution to the Bistritzer-MacDonald model.\cite{Bistritzer_2011} We employ the implementation for the TBG lattices in which the first layer is clockwise rotated by a half twist angle, $\theta_1 = -\theta/2$ and the second layer is counterclockwise rotated by $\theta_2 = +\theta/2$. The unrotated layers are defined by unit vectors $\vect{a}_1 = a.\vect{e}_x$ and $\vect{a}_2 = a\cos\pi/3.\vect{e}_x+a\sin\pi/3.\vect{e}_y$, where $a$ is the lattice constant of graphene, and $\vect{e}_x,\vect{e}_y$ the unit vectors defining the $x,y$ directions, respectively. The two associated vectors defining the reciprocal lattice are thus $\vect{a}_1^\star = (4\pi/a)(\cos\pi/6.\vect{e}_x+\sin\pi/6.\vect{e}_y)$ and $\vect{a}_2^\star = (4\pi/a)\vect{e}_y$. Under the twisting these vectors defining the atomic lattice of the two graphene layers as well as their reciprocal lattice are determined by applying the rotation matrix $R_z(\theta_\ell)$, specifically, $\vect{a}_i^{(\ell)} = R_z(\theta_\ell)\cdot\vect{a}_i$, and  $\vect{a}_i^{\star(\ell)} = R_z(\theta_\ell)\cdot\vect{a}_i^{\star}$. The first Brillouin zone $\text{BZ}^{(\ell)}$ of each graphene layer is defined by a set of six corner points $\vect{K}_i^{(\ell)}$. For instance, the point $\vect{K}_{1}^{(\ell)}$ is determined by the point $\vect{K}_{-}^{(\ell)}$ with $\vect{K}_{\xi}^{(\ell)} = -\xi(2\vect{a}_1^{\star(\ell)}+\vect{a}_2^{\star(\ell)})/3, \xi = \pm 1$.

For the commensurate twisting, the twist angle $\theta$ is determined by two integer numbers $m,n$ via the formula:
\begin{equation}\label{Eq12}
    \theta =  \arctan\left(\frac{|n^2-m^2|\sin(\pi/3)}{(n^2+m^2)\cos(\pi/3)+2mn}\right).
\end{equation}
When $|m-n| = 1$, the moire reciprocal lattice is defined by two unit vectors $\vect{A}_i^\star = \vect{a}_i^{\star(1)}-\vect{a}_i^{\star(2)}$. The mini-Brillouin zone is defined by six $\vect{K}^M_i$ points given by: $\vect{K}_1^M = (-\vect{A}_1^\star+\vect{A}_2^\star)/3$ and $\vect{K}^M_6 = (\vect{A}_1^\star+2\vect{A}_2^\star)/3$. The other points are found by shifting these points via the reciprocal unit vectors. 

Given a Hamiltonian $H$ for electrons, we first use the layer-resolution vector basis set $\{|\ell\rangle\,|\,\ell = 1,2\}$ to represent it. Using the identity $1=\sum_{\ell=1}^2|\ell\rangle\langle\ell|$ we have:
\begin{equation}\label{Eq13}
    H = \sum_{\ell,\ell^\prime=1}^2|\ell\rangle H_{\ell,\ell^\prime}\langle\ell^\prime|,
\end{equation}
where $H_{\ell,\ell^\prime} = \langle\ell|H|\ell^\prime\rangle$. Next, we use the lattice-resolution vector basis set $\{|\vect{k},m\rangle = |\vect{k}+\vect{G}_m\rangle \,|\,\vect{k}\in BZ, m\in \mathbb{Z}\}$ to further specify $H_{\ell,\ell^\prime}$. Also using the identity $1=\sum_{\vect{k}\in BZ,m}|\vect{k},m\rangle\langle\vect{k},m|$ we have:
\begin{align}\label{Eq14}
    H 
    &= \sum_{\vect{k}\in BZ}\sum_{\ell,m}\sum_{\ell^\prime,n}|\ell,\vect{k},m\rangle H_{\ell,m;\ell^\prime,n}(\vect{k})\langle\ell^\prime,\vect{k},n|,
\end{align}
where we denote $|\ell,\vect{k},m\rangle = |\ell\rangle|\vect{k},m\rangle$ and use the relation $ \langle\ell,\vect{k},m|H|\ell^\prime,\vect{k}^\prime,n\rangle = H_{\ell,m;\ell^\prime,n}(\vect{k})\delta_{\vect{k},\vect{k}^\prime}$. Accordingly, the low energy states of electrons in the TBG configurations of tiny twist angles are distinguished by a quantum index $\xi = \pm1$ that corresponds to the two nonequivalent Dirac valleys $\vect{K}_{\xi}^{(1,2)}$ of the graphene mono-layers.\cite{Bistritzer_2011} Notice that as the tiny twist angles, the position of the two points $\vect{K}_{\xi}^{(1)}$ and $\vect{K}_{\xi}^{(2)}$ are close to each other. The  Bistritzer-MacDonald model is given by a Hamiltonian in the real-space representation as follows:\cite{Bistritzer_2011}
\begin{align}\label{Eq15}
    \hat{H}^\xi = 
    \left(\begin{array}{cc}
        H_1^\xi(\hat{\vect{p}}) & T^\xi(\hat{\vect{r}}) \\
        T^{\xi\dagger}(\hat{\vect{r}}) & H_2^\xi(\hat{\vect{p}})
    \end{array}\right).
\end{align}
Here $\hat{\vect{p}}$ and $\hat{\vect{r}}$ are the momentum and position operators, respectively. The Hamiltonians $H_1^\xi(\hat{\vect{p}})$ and $H_2^\xi(\hat{\vect{p}})$ of the two uncoupled graphene layers are given by the 2D Dirac Hamiltonian:
\begin{equation}\label{Eq16}
    H_\ell^\xi(\hat{\vect{p}}) = - v_F(\xi\sigma_x,\sigma_y)\cdot\left[R^z\left(-\theta_\ell\right)\cdot(\hat{\vect{p}}-\hbar\vect{K}_\xi^\ell)\right],
\end{equation}
where $\theta_\ell = (-1)^{\ell}\theta/2$ is the rotation angle of layer $\ell$ with respect to the fixed Cartesian coordinate axes $Oxyz$;  $R^z\left(-\theta_\ell\right)$ is a matrix to rotate back the relevant vectors around the $Oz$ axis to keep the canonical form of the Dirac Hamiltonian; $\vect{K}_\xi^\ell$ is the corner point of type (valley) $\xi$ of the first Brillouin zone of the layer $\ell$; $\sigma_x,\sigma_y$ are two conventional Pauli matrices; and $v_F$ is the Fermi velocity (The minus sign is due to the negative value of the hopping parameter, $V_{pp\pi} = -2.7$ eV). The interlayer coupling block term is given by:\cite{Koshino_2018}
\begin{align}\label{Eq17}
    T^\xi(\hat{\vect{r}}) =& \left(\begin{array}{cc}
    u & u^\prime \\
    u^\prime    & u 
    \end{array}\right)+\left(\begin{array}{cc}
    u & u^\prime\omega^{-\xi} \\
    u^\prime\omega^{\xi}    & u 
    \end{array}\right)e^{i\xi\delta\vect{k}_2\cdot\hat{\vect{r}}}\nonumber\\
    &+\left(\begin{array}{cc}
    u & u^\prime\omega^{\xi} \\
    u^\prime\omega^{-\xi}    & u 
    \end{array}\right)e^{i\xi\delta\vect{k}_3\cdot\hat{\vect{r}}},
\end{align}
where $\omega = \exp(i2\pi/3)$ and $\delta\vect{k}_2 = \vect{A}_1^\star$ and $\delta\vect{k}_3 = \vect{A}_1^\star+\vect{A}_2^\star$. We choose the value of two parameters $u = 0.0797$ eV and $u^\prime = 0.0975$ eV to take into account of effects of lattice reconstruction.\cite{Koshino_2018}

Associated with the Hamiltonian (\ref{Eq16}), the velocity operator is determined by:
\begin{equation}\label{Eq18}
    \hat{v}_\alpha = -v_F(\xi\sigma_x,\sigma_y)\cdot R^z_{\alpha}(-\theta_\ell),
\end{equation}
where $R^z_{\alpha}(-\theta_\ell)$ denotes the first/second column of the rotation matrix $R^z(-\theta_\ell)$ if  $\alpha=x/y$. It should be noticed that because of the relativistic form of $H^\xi_\ell(\hat{\vect{p}})$ given by Eq. (\ref{Eq13}), the current operator $\hat{j}_\alpha$ has only one contribution part of drift current, $\hat{j}_\alpha = e\hat{v}_\alpha$, i.e., no diffusion and diamagnetic parts.

We find the spectrum of the Hamiltonian $\hat{H}$ by solving this secular equation:
\begin{align}\label{Eq19}
    \left(\begin{array}{cc}
        H_1^\xi(\hat{\vect{p}}) & T^\xi(\hat{\vect{r}}) \\
        T^{\xi\dagger}(\hat{\vect{r}}) & H_2^\xi(\hat{\vect{p}})
    \end{array}\right)|\psi_{\xi,\vect{k}}\rangle = E|\psi_{\xi,\vect{k}}\rangle.
\end{align}
Due to the ``approximate'' periodicity of the moire lattice of TBG lattices, electron states are determined as the Bloch state vectors $|\psi_{\xi,\vect{k}}\rangle$ that are expanded in terms of plane-wave vectors as follows:
\begin{equation}\label{Eq20}
    |\psi_{\xi,\vect{k}}\rangle = \sum_m\left(\begin{array}{c}
         C_{1,\xi,\vect{k}}(\vect{G}_m)  \\
         C_{2,\xi,\vect{k}}(\vect{G}_m)
    \end{array}\right)|\vect{k}+\vect{G}_m\rangle,
\end{equation}
where $\hat{\vect{p}}|\vect{k}+\vect{G}_m\rangle = (\vect{k}+\vect{G}_n)|\vect{k}+\vect{G}_n\rangle$,  $\langle\vect{r}|\vect{k}+\vect{G}_m\rangle = e^{i(\vect{k}+\vect{G}_m)\cdot\vect{r}}$ and $C_{1,\xi,\vect{k}}(\vect{G}_m)$ and $C_{2,\xi,\vect{k}}(\vect{G}_m)$ are the 2D vectors of  combination coefficients that are need to be found. Here $\{\vect{G}_n\,|\, n = 1,2,3, \hdots, N_{\vect{G}}\}$ is a set of $N_{\vect{G}}$ vectors of the moire reciprocal lattice. Substitute this trial expression into Eq. (\ref{Eq19}) and left-multiplying both sides with $\langle\vect{k}+\vect{G}_n|$,  we have:
\begin{align}\label{Eq21}
    &\sum_m\langle\vect{k}+\vect{G}_n|\hat{H}^\xi|\vect{k}+\vect{G}_m\rangle\left(\begin{array}{c}
         C_{1,\xi,\vect{k}}(\vect{G}_m)  \\
         C_{2,\xi,\vect{k}}(\vect{G}_m)
    \end{array}\right) \nonumber\\
    &\hspace{3.2cm}= E\left(\begin{array}{c}
         C_{1,\xi,\vect{k}}(\vect{G}_n)  \\
         C_{2,\xi,\vect{k}}(\vect{G}_n).
    \end{array}\right)
\end{align}
With the notice that
\begin{subequations}
\begin{align}
    \langle\vect{k}+\vect{G}_n|H_\ell^\xi|\vect{k}+\vect{G}_m\rangle &= H_\ell^\xi(\vect{k}+\vect{G}_n)\delta_{\vect{G}_n,\vect{G}_m},\label{Eq22a}\\
    \langle\vect{k}+\vect{G}_n|T^\xi(\hat{\vect{r}})|\vect{k}+\vect{G}_m\rangle &= \sum_{j=1}^3T_j^\xi\delta_{\vect{G}_n,\vect{G}_m+\delta\vect{k}_j},\label{Eq22b}\\
    \langle\vect{k}+\vect{G}_n|T^{\xi\dagger}(\hat{\vect{r}})|\vect{k}+\vect{G}_m\rangle &= \sum_{j=1}^3T_j^{\xi\dagger}\delta_{\vect{G}_n,\vect{G}_m-\delta\vect{k}_j},\label{Eq22c}
\end{align}
\end{subequations}
we specify Eq. (\ref{Eq21}) in the form:
\begin{widetext}
\begin{align}\label{Eq23}
    &\sum_m\left[\left(\begin{array}{cc}
    H_1^\xi(\vect{k}+\vect{G}_n)     & T_1^\xi \\
    T_1^{\xi\dagger}     & H_2^\xi(\vect{k}+\vect{G}_n)
    \end{array}\right)\delta_{\vect{G}_n,\vect{G}_m}+\right.\nonumber\\
    &\hspace{2cm}+\left.\sum_{j=2}^3\left(\begin{array}{cc}
    0  & T_j^\xi \\
    0  & 0
    \end{array}\right)\delta_{\vect{G}_n,\vect{G}_m+\delta\vect{k}_j}+\sum_{j=2}^3\left(\begin{array}{cc}
    0  & 0 \\
    T_j^{\xi\dagger}  & 0
    \end{array}\right)\delta_{\vect{G}_n,\vect{G}_m-\delta\vect{k}_j}\right]\left(\begin{array}{c}
         C_{1,\xi,\vect{k}}(\vect{G}_m)  \\
         C_{2,\xi,\vect{k}}(\vect{G}_m)
    \end{array}\right) = E\left(\begin{array}{c}
         C_{1,\xi,\vect{k}}(\vect{G}_n)  \\
         C_{2,\xi,\vect{k}}(\vect{G}_n)
    \end{array}\right).
\end{align}
\end{widetext}
With a set of reciprocal lattice vectors $\{\vect{G}_n\,|\,n = 1,2,3,\hdots, N_{\vect{G}}\}$, the above equation is the representative of a set of $N_G$ linear equations for the coefficients $(C_{1,\xi,\vect{k}}(\vect{G}_n),C_{2,\xi,\vect{k}}(\vect{G}_n))^T$. Numerically, for each value of $\xi$ and each value of $\vect{k}\in \text{MBZ}$, we define a $4N_G\times 4N_G$ Hermitian matrix $H_{\vect{k}}^\xi$ that are given in the blocks as follows:
\begin{subequations}
\begin{align}
    &[H_{\vect{k}}^\xi]_{n,n} = \left(\begin{array}{cc}
    H_1^\xi(\vect{k}+\vect{G}_n)     & T_1^\xi \\
    T_1^{\xi\dagger}     & H_2^\xi(\vect{k}+\vect{G}_n)
    \end{array}\right),\label{Eq24a}\\
    &[H_{\vect{k}}^\xi]_{n,m_j} = \left(\begin{array}{cc}
    0     & T_j^\xi \\
    0     & 0
    \end{array}\right), \hspace{0.5cm}\text{if}\hspace{0.5cm} \vect{G}_{m_j} = \vect{G}_n-\delta\vect{k}_j,\label{Eq24b}\\
    &[H_{\vect{k}}^\xi]_{n,m_j^\prime} = \left(\begin{array}{cc}
    0     & 0 \\
    T_j^{\xi\dagger}     & 0
    \end{array}\right), \hspace{0.4cm}\text{if}\hspace{0.5cm} \vect{G}_{m_j^\prime} = \vect{G}_n+\delta\vect{k}_j.\label{Eq24c}
\end{align}
\end{subequations}
Diagonalize the matrix $H_{\vect{k}}^\xi$, we obtain all possible eigen-values of $E_n^{\xi}(\vect{k})$. From these data, we can represent the electronic energy band structure. The issue here is the value of $N_{\vect{G}}$. Since the model is valid for low energy range in which the energy surfaces of the monolayer graphene take the cone geometry, we thus define a cutoff energy $E_c$, then determine $N_{\vect{G}}$ the number of $\vect{G}_n$ vectors such that $\|\vect{G}_n\|\leq E_c/\hbar v_F$.
\begin{figure*}\centering
\includegraphics[clip=true,trim=1.7cm 7cm 7cm 7cm,width=0.4\textwidth]{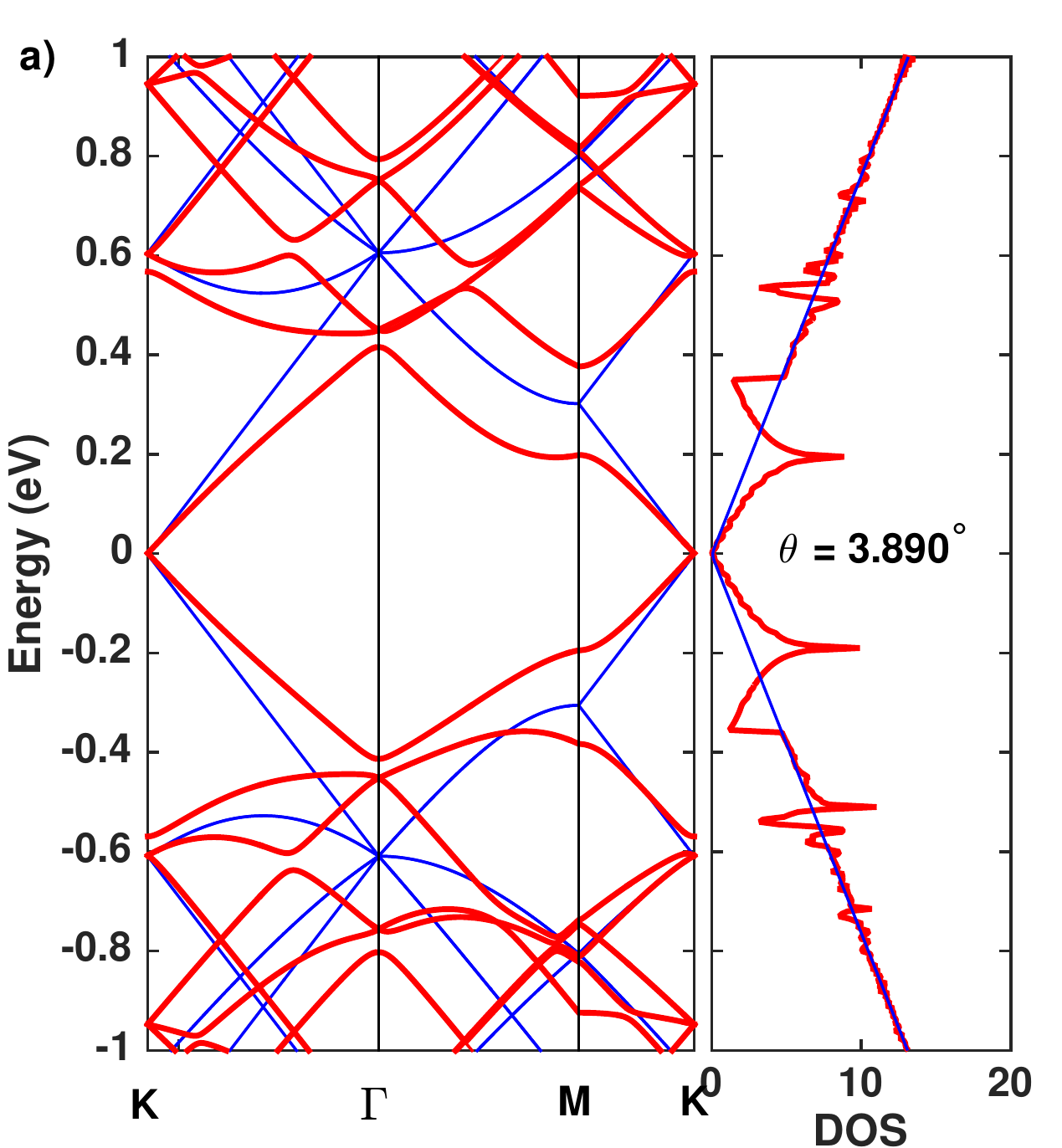}
\includegraphics[clip=true,trim=1.7cm 7cm 11.36cm 7cm,width=0.265\textwidth]{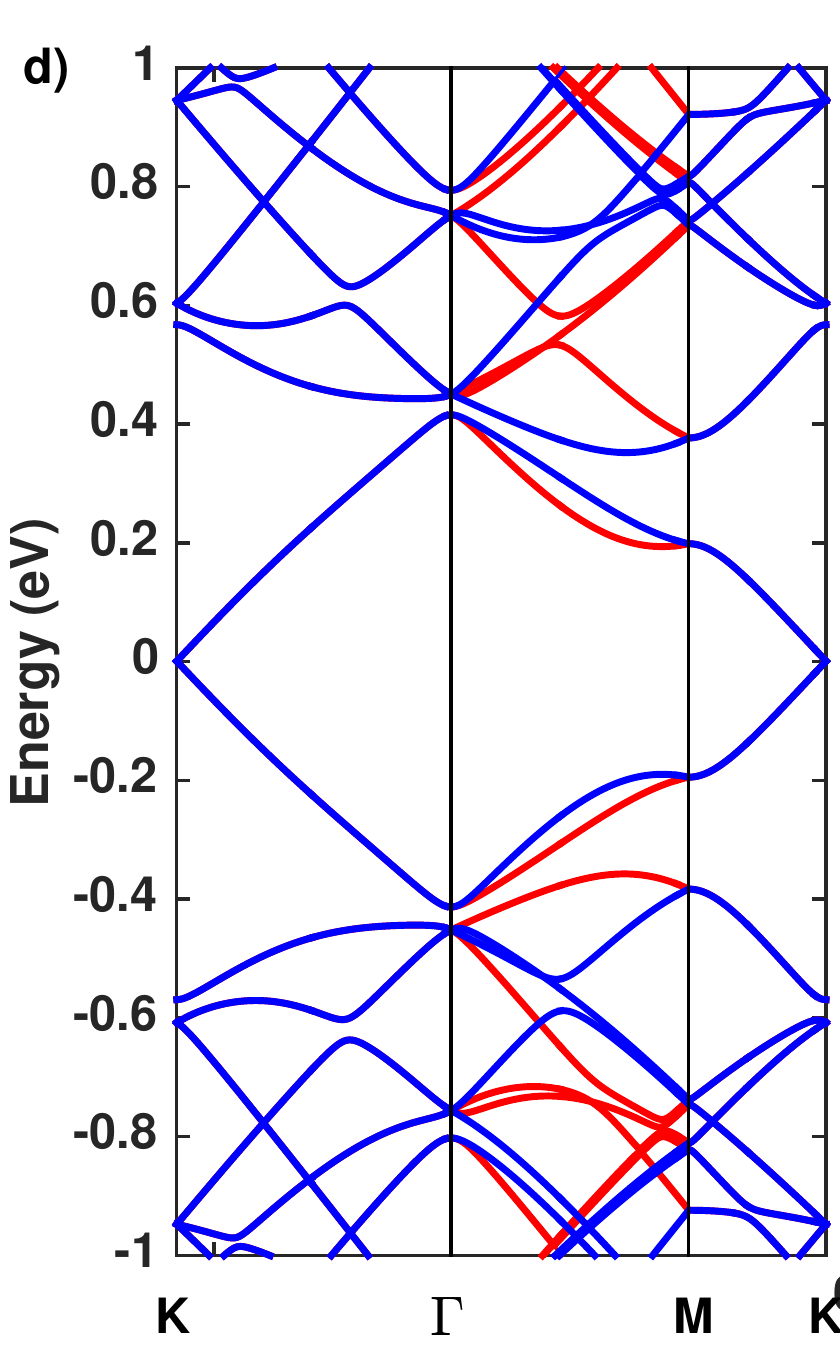}\\
\includegraphics[clip=true,trim=1.7cm 7cm 7cm 7cm,width=0.4\textwidth]{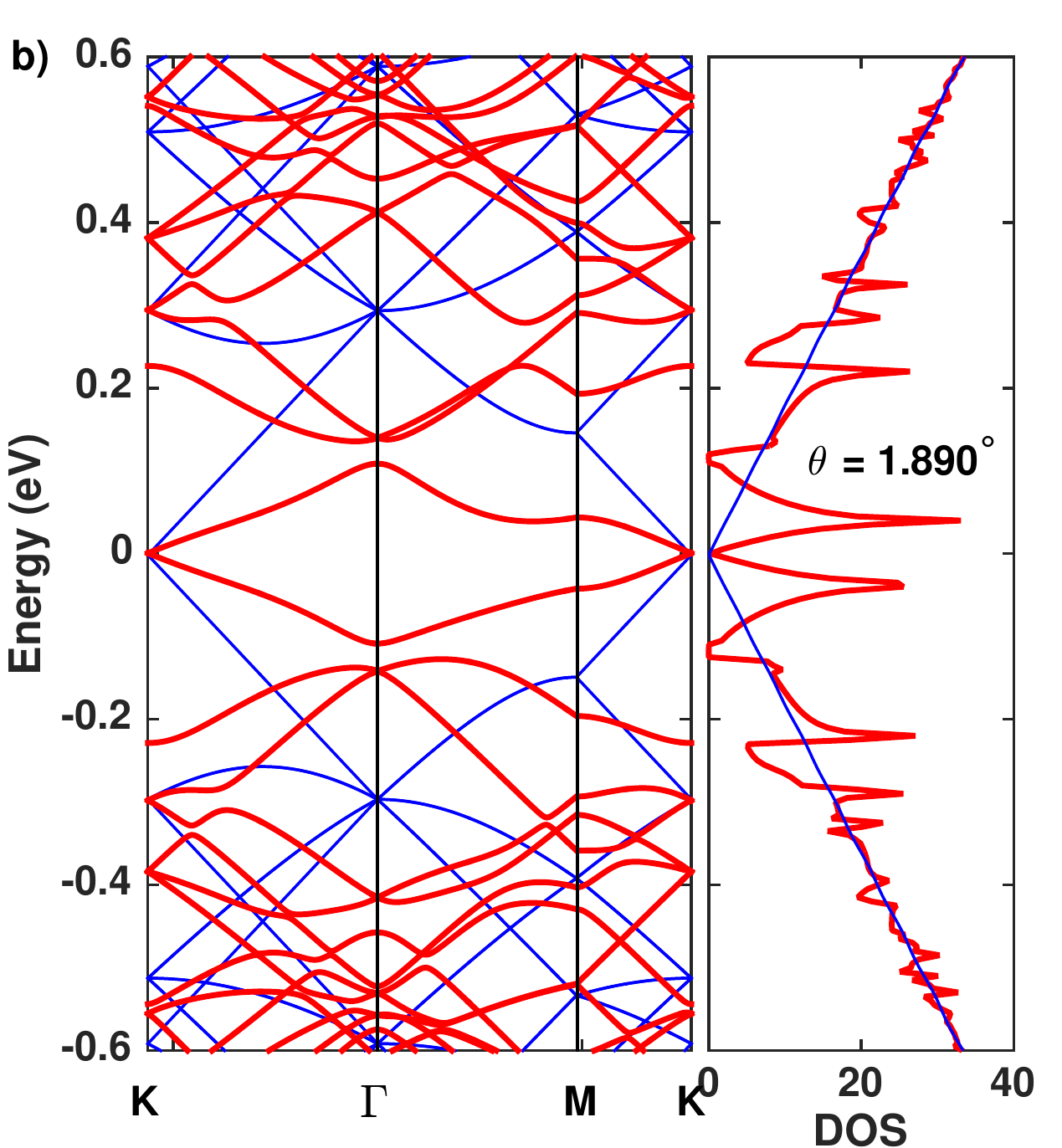}
\includegraphics[clip=true,trim=1.7cm 7cm 11.36cm 7cm,width=0.265\textwidth]{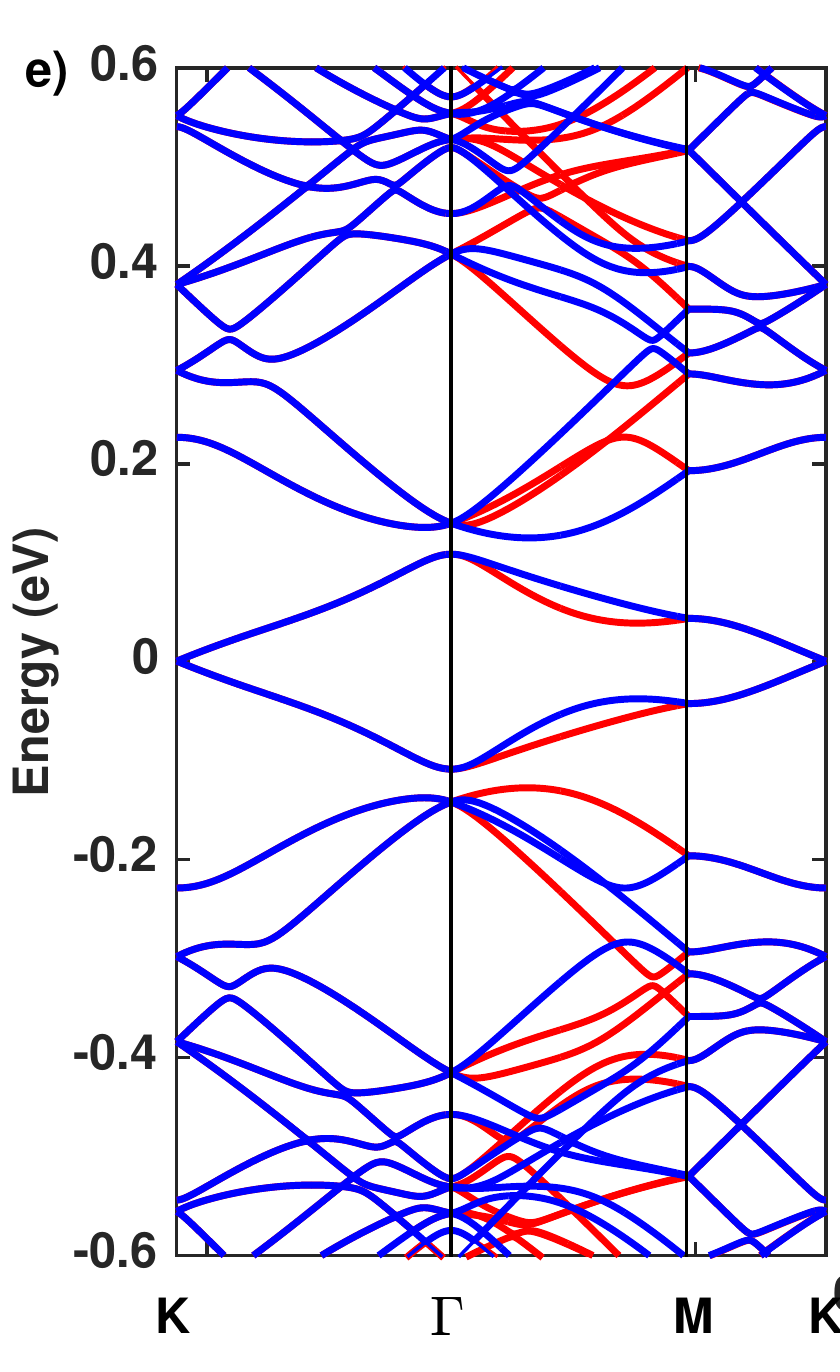}\\
\includegraphics[clip=true,trim=1.7cm 7cm 7cm 7cm,width=0.4\textwidth]{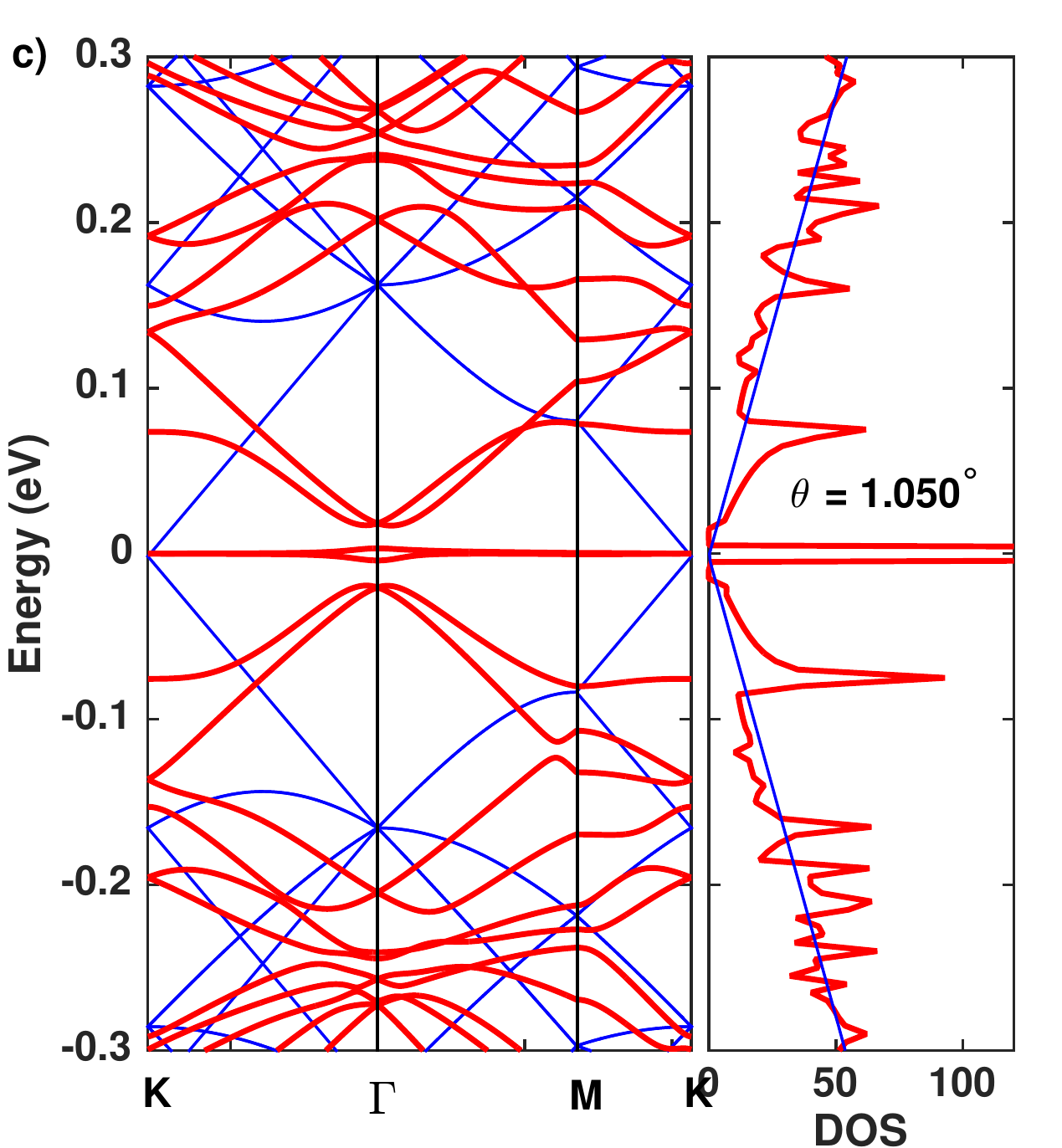}
\includegraphics[clip=true,trim=1.7cm 7cm 11.36cm 7cm,width=0.265\textwidth]{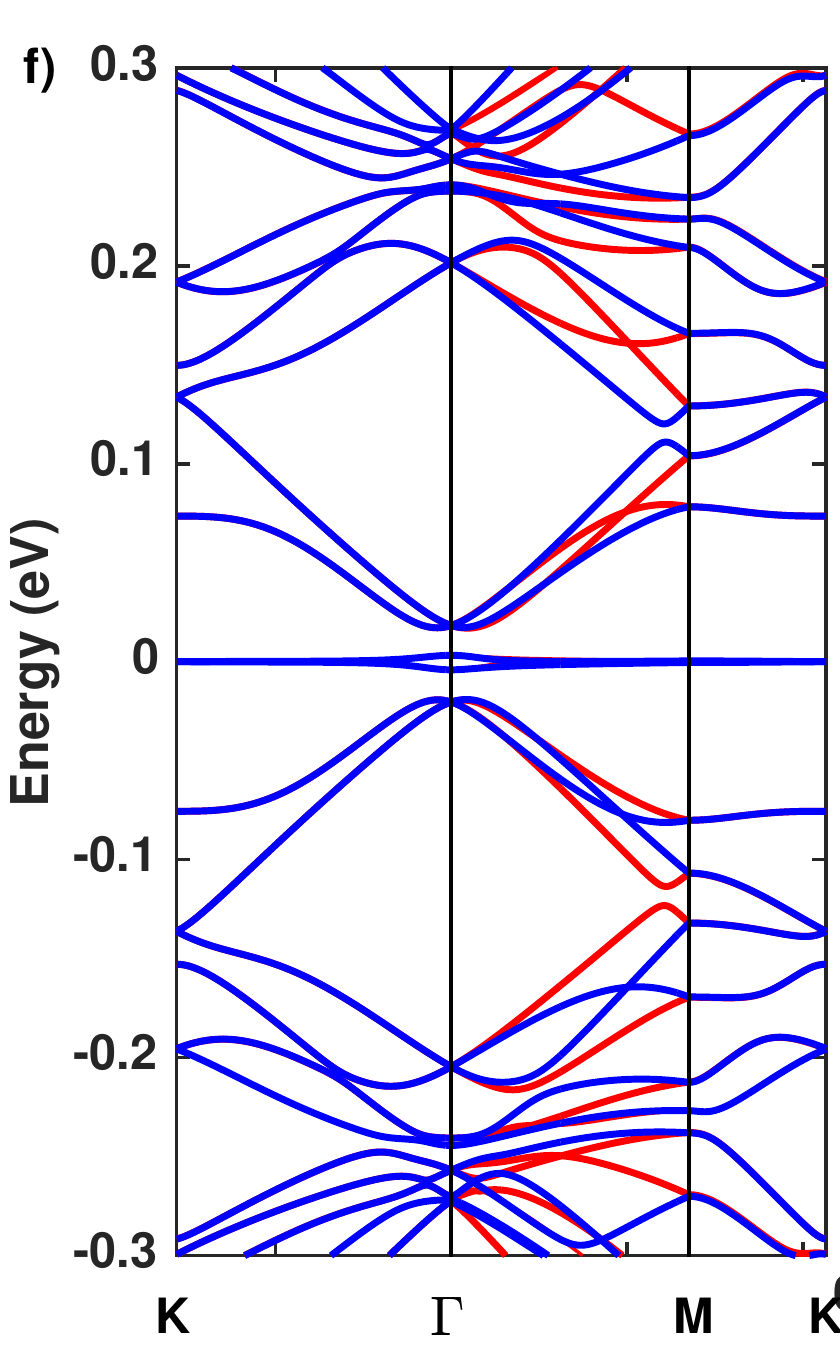}
\caption{\label{Fig1} These present the energy band structures of electron in three TBG configurations with $\theta = 3.890^\circ$ (a,d), $1.890^\circ$ (b,e), and $1.050^\circ$ (c,f). The blue curves in (a,b,c) are for the dispersion and density of states of graphene, which are obtained by turning off the interlayer coupling. The blue and red curves in (d,e,f) are the dispersion curves obtained for the valley $\xi = +1$ and $\xi = -1$.}
\end{figure*}
\section{Results and discussions}
\subsection{Electronic band structure of TBGs}
We employed the numerical method to solve Eq. (\ref{Eq23}). Specifically, for each value of $\xi=\pm 1$, a Hermitian matrix  $H_{\vect{k}}^\xi$ of size $4N_G\times 4N_G$ was constructed and then diagonalized for each value of $\vect{k}$ given in the mini-Brillouin zone of TBG systems. The sets of eigen-values $\{E_m(\vect{k})\}$ and of eigen-vectors $\{\mathbf{C}^m_{\xi}(\vect{k})=[C^m_{1,\xi,\vect{k}}(\vect{G}_1),C^m_{2,\xi,\vect{k}}(\vect{G}_1),\hdots,C^m_{1,\xi,\vect{k}}(\vect{G}_{N_G}),C^m_{2,\xi,\vect{k}}(\vect{G}_{N_G})]^T\}$ were determined. Obtained results for three TBG configurations with the twist angles of $\theta = 1.05^\circ, 1.89^\circ$ and $3.89^\circ$ were presented in Fig. \ref{Fig1}. The blue curves are the results obtained as we artificially turn off the electronic interlayer coupling. We plot both data on the same figure for the aim of comparison, i.e., in order to highlight the effects of the interlayer coupling. The appearance of many dispersion curves in the figure is the result of the folding of the energy band structure of two graphene layers due to the enlargement of the unit cell of the TBG lattices, hence the shrink of the Brillouin zone into the mini-Brillouin zone. However, the electronic band structure of the TBG systems is not simply formed like that. At the points of crossing bands of monolayer graphene, the interlayer coupling leads to the hybridization of Bloch states; the energy degeneracy is thus lifted to form the new bands. Our obtained results are in agreement with other available data in the literature.\cite{Morell_2010,Laissardiere_2012,Moon-2013,Koshino_2018} However, in this work, we would like to emphasize the formation of new electronic states which are unique in the TBG systems because they govern the electronic, optical and transport properties. Fig. \ref{Fig1} can be seen as the evolution picture of the band structure with respect to the twist angle. We realize that in the low energy range around the intrinsic Fermi energy ($E_F= 0$), the dispersion curves of the configuration with $\theta = 3.89^\circ$ qualitatively have the similar form of the blue curves although they are not identical. This implies that the created states in the bilayer system much share the essence of states in monolayer graphene. For smaller twist angles, in particular the one with $\theta = 1.05^\circ$, a band around the Fermi energy with a very narrow bandwidth is formed and separated from the lower and upper bands by a narrow gap. These dispersion curves are totally different from any blue curves, see Fig. \ref{Fig1}(e). We therefore conclude that the electronic interlayer coupling creates unique states in the TBG lattices of tiny twist angles. Although the TBG systems are two-dimensional, geometrical features of the energy surfaces in the wave vector space are difficult to be directly analyzed. However, general features should exhibit in the picture of density of states (DOS) as the van Hove singularity behaviors shown in the right panels of Figs. \ref{Fig1}(a,b,c). In these figures, the blue curves are numerically obtained without taking into account of the electronic interlayer coupling, and thus the DOS of two independent graphene layers. The red curves show significant peaks due to the van Hove singularities, reflecting the existence of extremal points and saddle points of the energy surfaces. Another important point in the TBG electronic structure is that the energy bands with the index $\xi +1$ (resulted from the hybridization of graphene Bloch states with the wave vector in the $K_1$ valley of the first layer and those in the $K_2$ valley of the second layer) is different from those with the index $\xi =-1$ (resulted from hybridization of Bloch states in the $K_1^\prime$ valley and in the $K_2^\prime$). The difference clearly manifests along the ${\it\Gamma}M$ direction as shown in Figs. \ref{Fig1}(d,e,f). However, it does not manifest in the picture of the density of states. We computed the DOS for $\xi = +1$ and $\xi = -1$ separately and obtained the identical results. This point implies that the energy surfaces with $\xi = +1$ are simply the rotation of the corresponding ones with $\xi = -1$ by an angle of $\pi/3$.

\begin{figure}\centering
\includegraphics[clip=true,trim=1.5cm 6.5cm 2cm 7cm,width=\columnwidth]{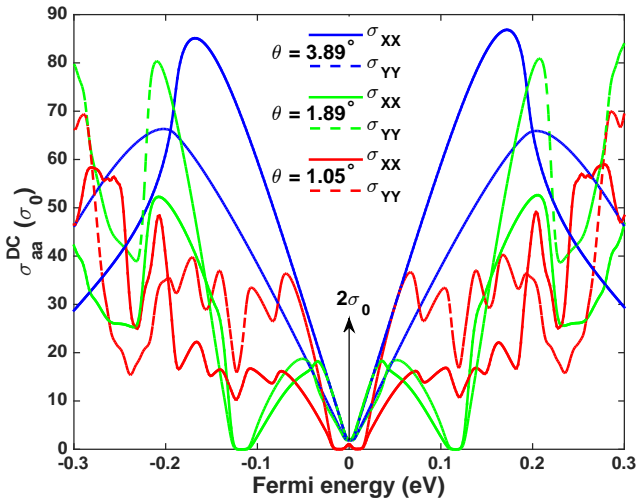}
\caption{\label{Fig2} These present the longitudinal DC conductivities $\sigma_{xx}^{DC}$ (solid curves) and $\sigma_{yy}^{DC}$ (dashed curves) of three TBG configurations with $\theta = 3.890^\circ$ (blue curves), $1.890^\circ$ (green curves) and $1.050^\circ$ (red curves).}
\end{figure}

\subsection{DC conductivity}
The calculation of the electronic structures of TBG configurations with not too small twist angles ($>2^\circ$) shows the appearance of two energy bands around the Fermi energy $E_F=0$ with the dispersion curves very similar to those of monolayer graphene, except the value of the Fermi velocity characterizing the linear dispersion law $v_F^{TBG} < v_F^{MLG}$. Meanwhile, if continuing to decrease the twist angle, these two bands shall become closer to each other and form an isolated energy band with a very narrow bandwidth. Such a ``flat-band'' at the Fermi energy was supposed not to contribute to the transport properties of TBG sheets, because of very small value of the Fermi velocity of carriers. In other words, electrons occupying states corresponding to these bands should be spatially confined in the atomic lattice. Actually, these bands are not totally dispersionless. Hence, it is really intriguing to investigate the transport properties of the TBG systems. In the framework of non-interacting electrons approximation, we calculated the DC conductivity using the  Kubo-Greenwood formula given by Eqs. (\ref{Eq9}) and (\ref{Eq10}). The results we have obtained for the longitudinal components $\sigma_{xx}^{DC}$ and $\sigma_{yy}^{DC}$ of three TBG configurations under study in the regime of low temperature are presented in Fig. \ref{Fig2}.

The data show that the ability of electric conducting of the TBG systems is anisotropic. This is quantified by a conductivity tensor with two different longitudinal elements $\sigma_{xx}^{DC}$ (the solid curves) and $\sigma_{yy}^{DC}$ (the dashed curves). In principle, the electronic structure must exhibit in the picture of the conductivity with respect to the Fermi energy. We calculated these two quantities for the two cases of $\xi = \pm 1$ and obtained the same value. This point is similar to the DOS picture. However, the difference of two longitudinal conductivities is the reflection of the anisotropic nature of the energy surfaces. For the TBG configuration with $\theta = 3.890^\circ$, the electronic structure is qualitatively similar to that of monolayer graphene (except the value of the Fermi velocity), therefore the conductivity curves have the typical V-shape (or the M-shape if considered in a larger energy range, see the blue curves). For the configurations with tiny twist angles, the picture of the electronic structure, and thus the conductivity picture, is much more complicated. However, our calculation shows that the conductivity at the intrinsic Fermi level $E_F =0$ always takes a finite non-zero value. Interestingly, this value, for the TBG configurations whose twist angles are not the magic ones, is about two times of $\sigma^{DC}_0 = 4e^2/\pi h$, a quantum value of the minimal conductivity of monolayer graphene. For the TBG configuration of the magic twist angle $\theta = 1.050^\circ$,  the conductivity curve exhibits a small hump at $E_F = 0$ with  $\sigma_{\alpha\alpha}^{DC}(0) \approx \sigma^{DC}_0$. The quantum value of graphene conductivity was a topic intensively discussed around 2006.\cite{Novoselov_2005,Katsnelson_2006,Tworzydlo_2006,Miao_2007,Blake_2009,Do_2010} Its origin lies in the fact that the valence and conduction bands are not separated, but elegantly touch each other at the $K$ points, so there are always available free carriers in graphene due to the inevitable fluctuations of the electrostatic profile around the Fermi energy level. Numerically, in our calculation, we approximated the delta-Dirac functions in the Kubo-Greenwood formula by a Gaussian function with the parameter $\eta$ that  defines the finite width of the peak. The value of this parameter was taken to be $\eta < 5$ meV. The use of $\eta$ is due to not only the reason for numerical calculation, but also the fact of the broadening of energy levels, and thus the finite lifetime of quasi-particles carrying the electric charge. The quantum value of the DC conductivity observed here for the TBG systems should have the same reason. However, it should be emphasized that it is supported by typical hybridized states in the TBG lattices.

\begin{figure}\centering
\includegraphics[clip=true,trim=2cm 11.5cm 5cm 7cm,width=\columnwidth]{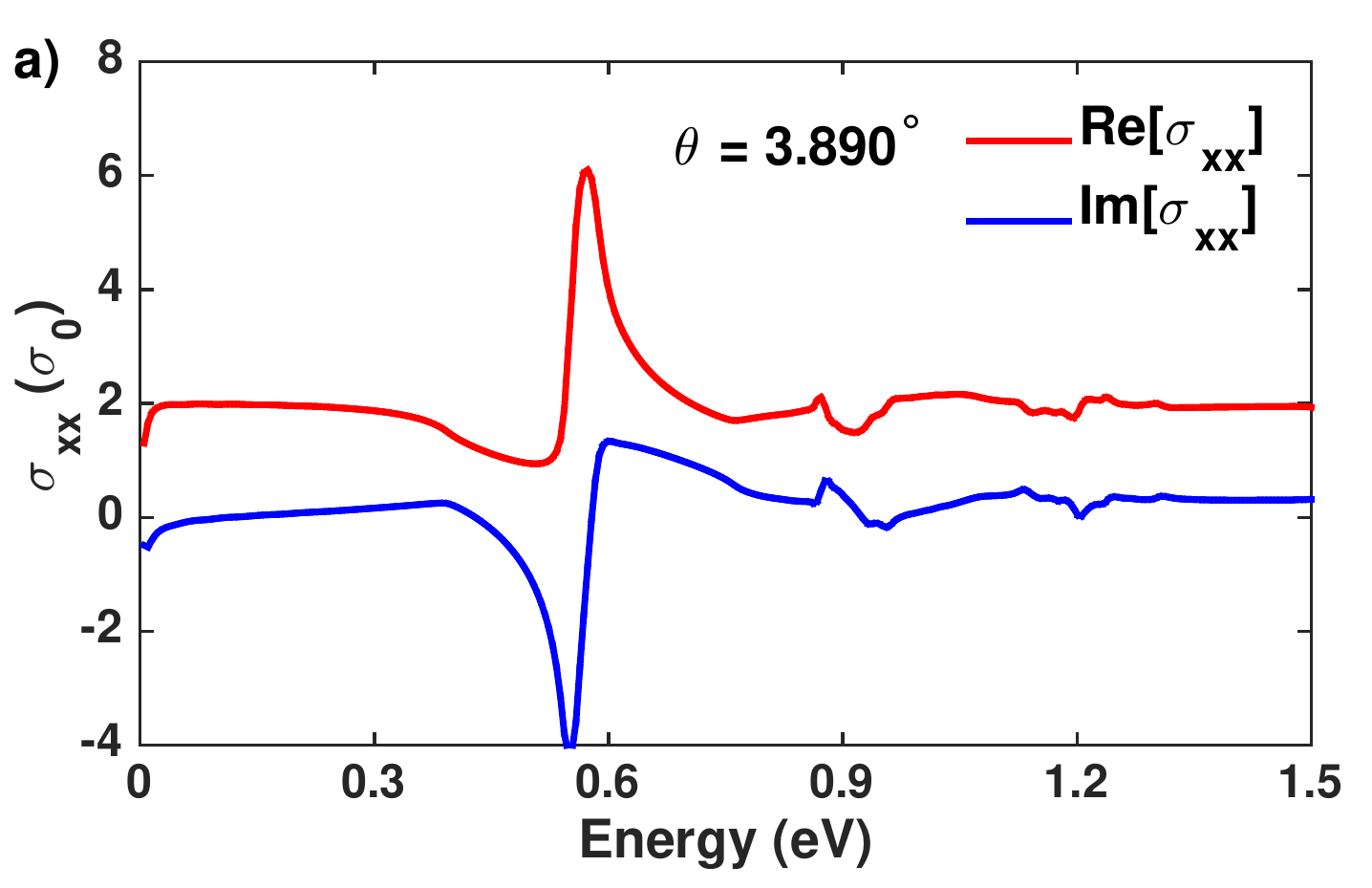}\\
\includegraphics[clip=true,trim=2cm 11.5cm 5cm 7cm,width=\columnwidth]{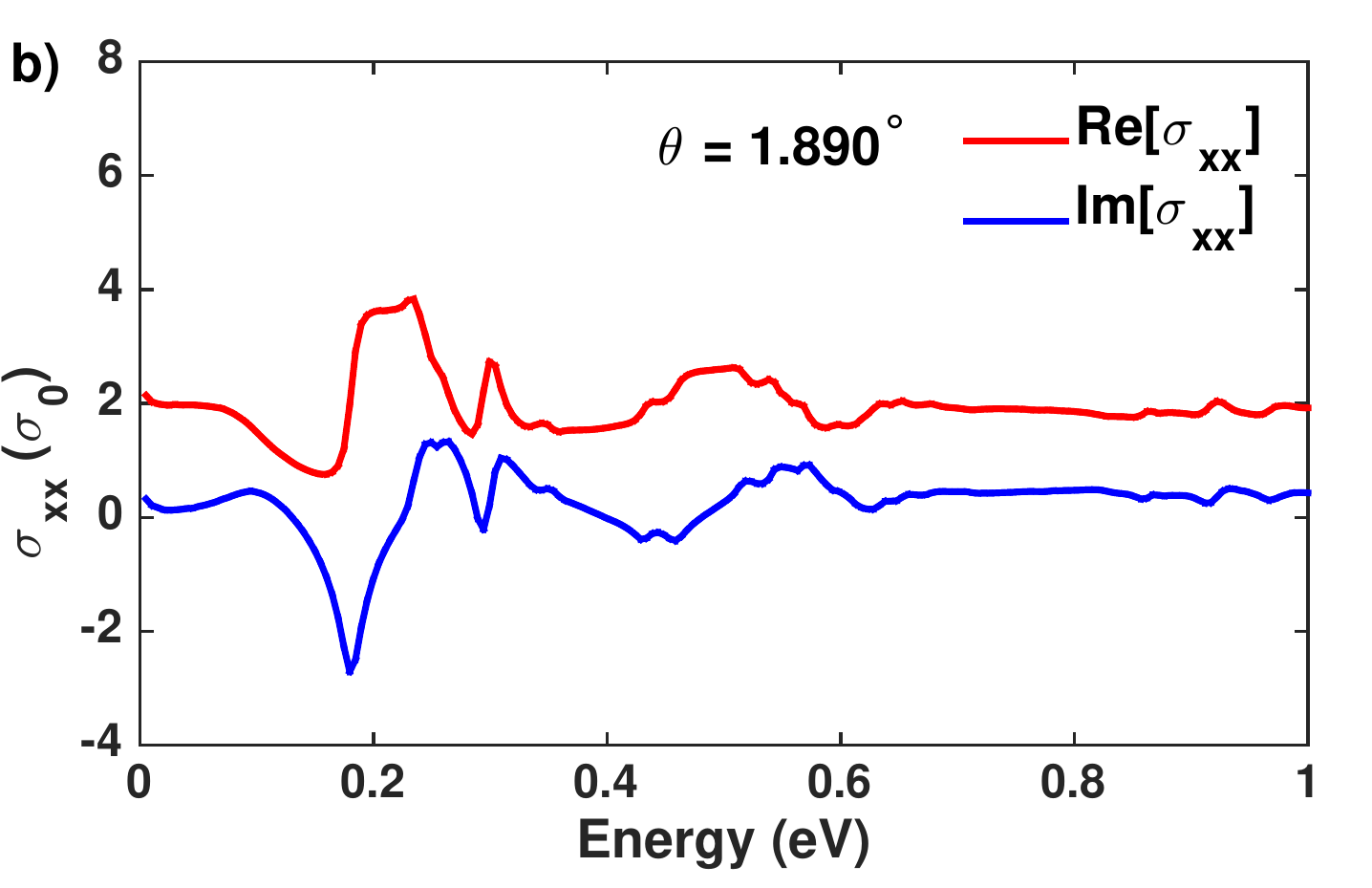}\\
\includegraphics[clip=true,trim=2cm 11.5cm 5cm 7cm,width=\columnwidth]{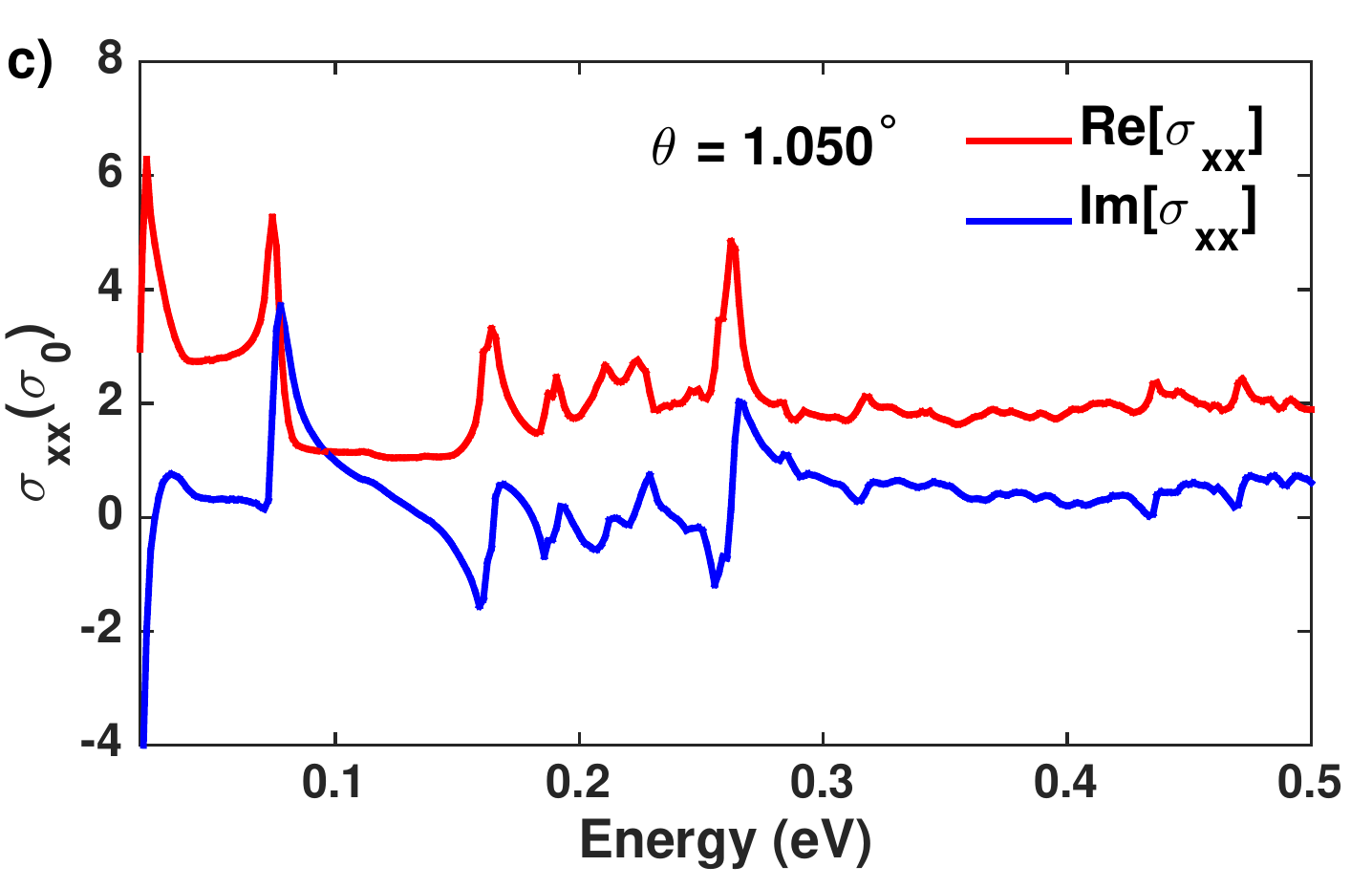}
\caption{\label{Fig3} These present the real and imaginary parts of the total longitudinal conductivity of the TBG configurations with $\theta = 3.890^\circ$ (a), $1.890^\circ$ (b) and $1.050^\circ$ (c). There are no total transverse conductivities owing to time-reversal symmetry.}
\end{figure}

\begin{figure}\centering
\includegraphics[clip=true,trim=1.6cm 11.5cm 5cm 7cm,width=\columnwidth]{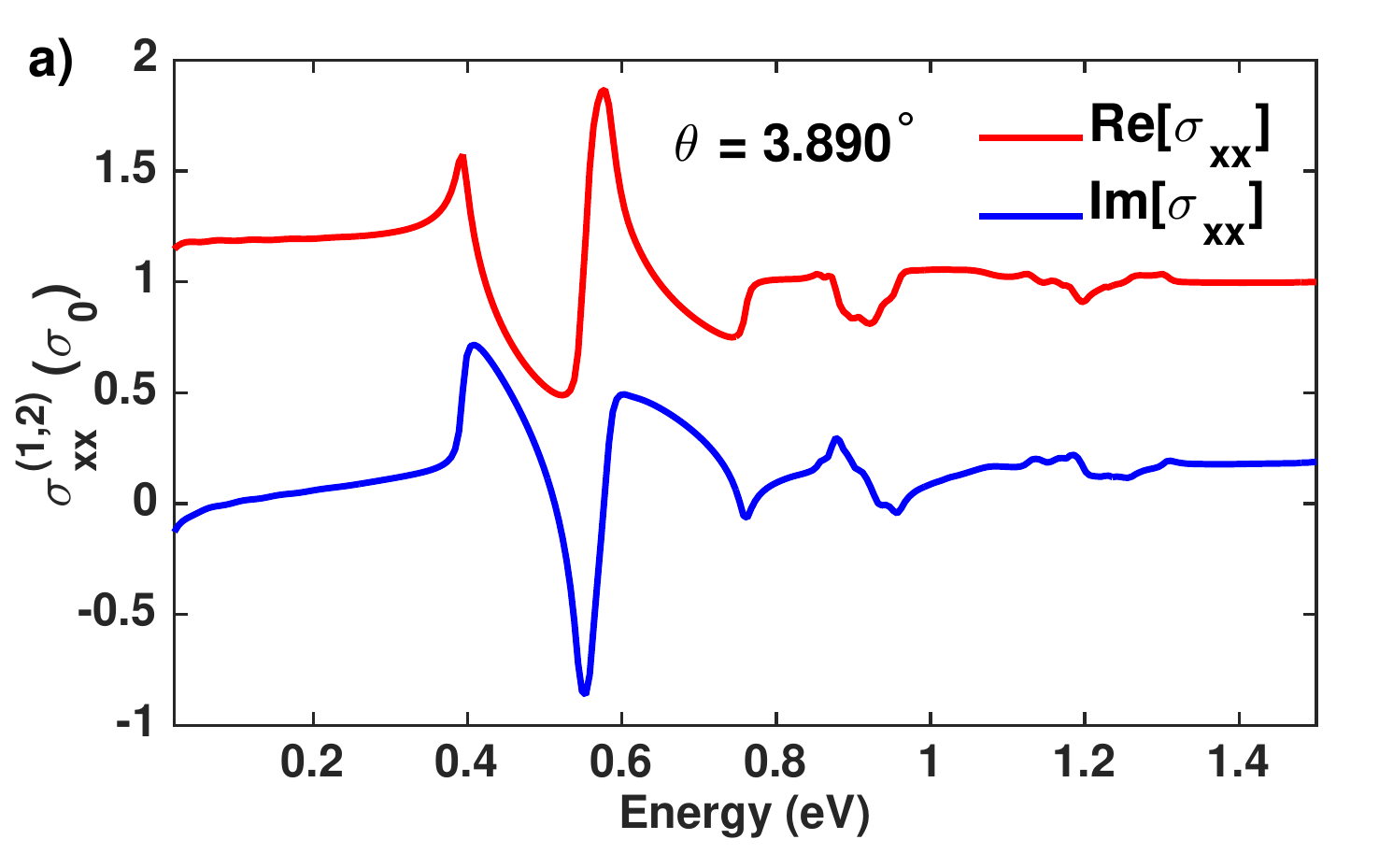}\\
\includegraphics[clip=true,trim=1.6cm 11.5cm 5cm 7cm,width=\columnwidth]{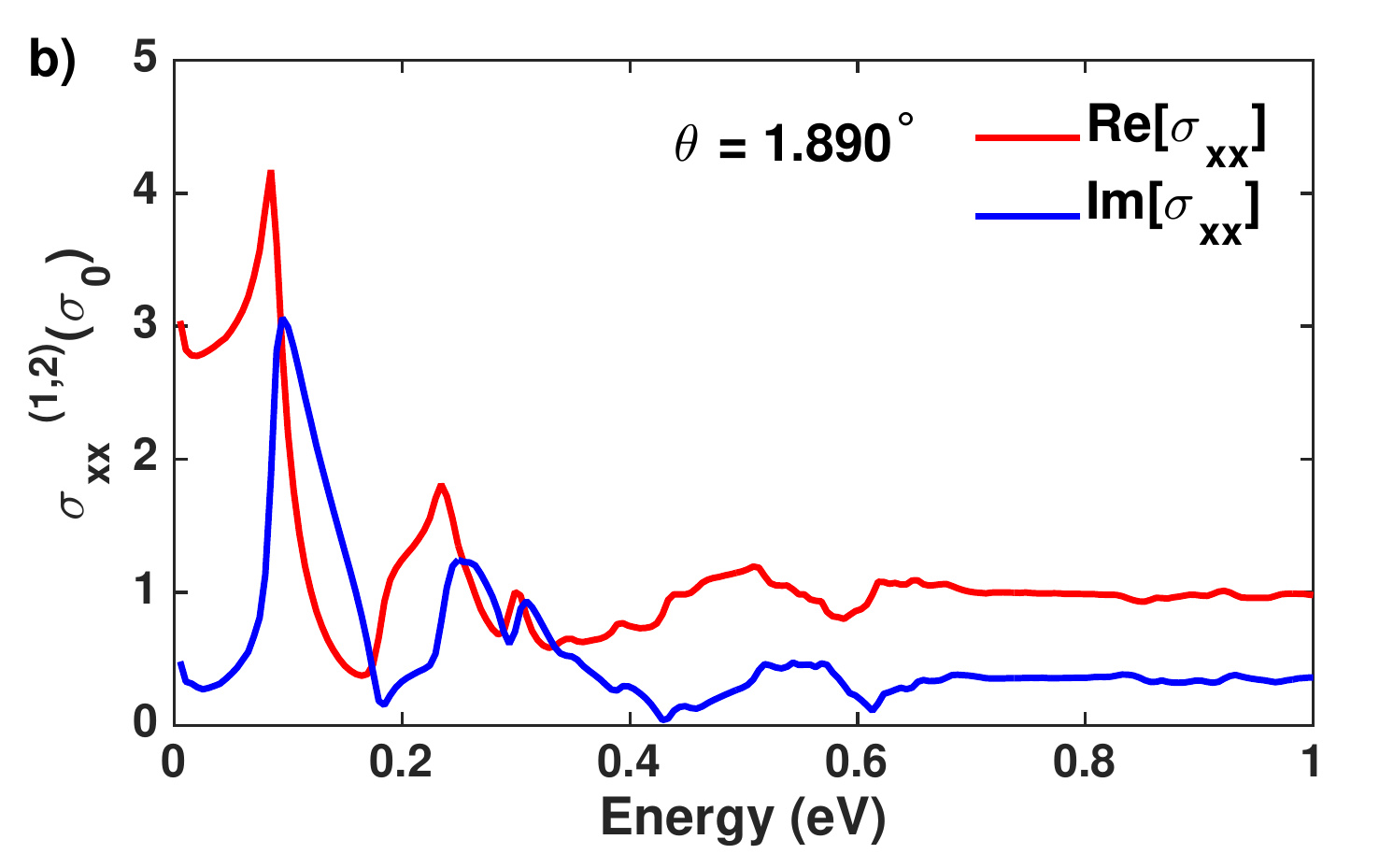}
\\
\includegraphics[clip=true,trim=1.6cm 11.5cm 5cm 7cm,width=\columnwidth]{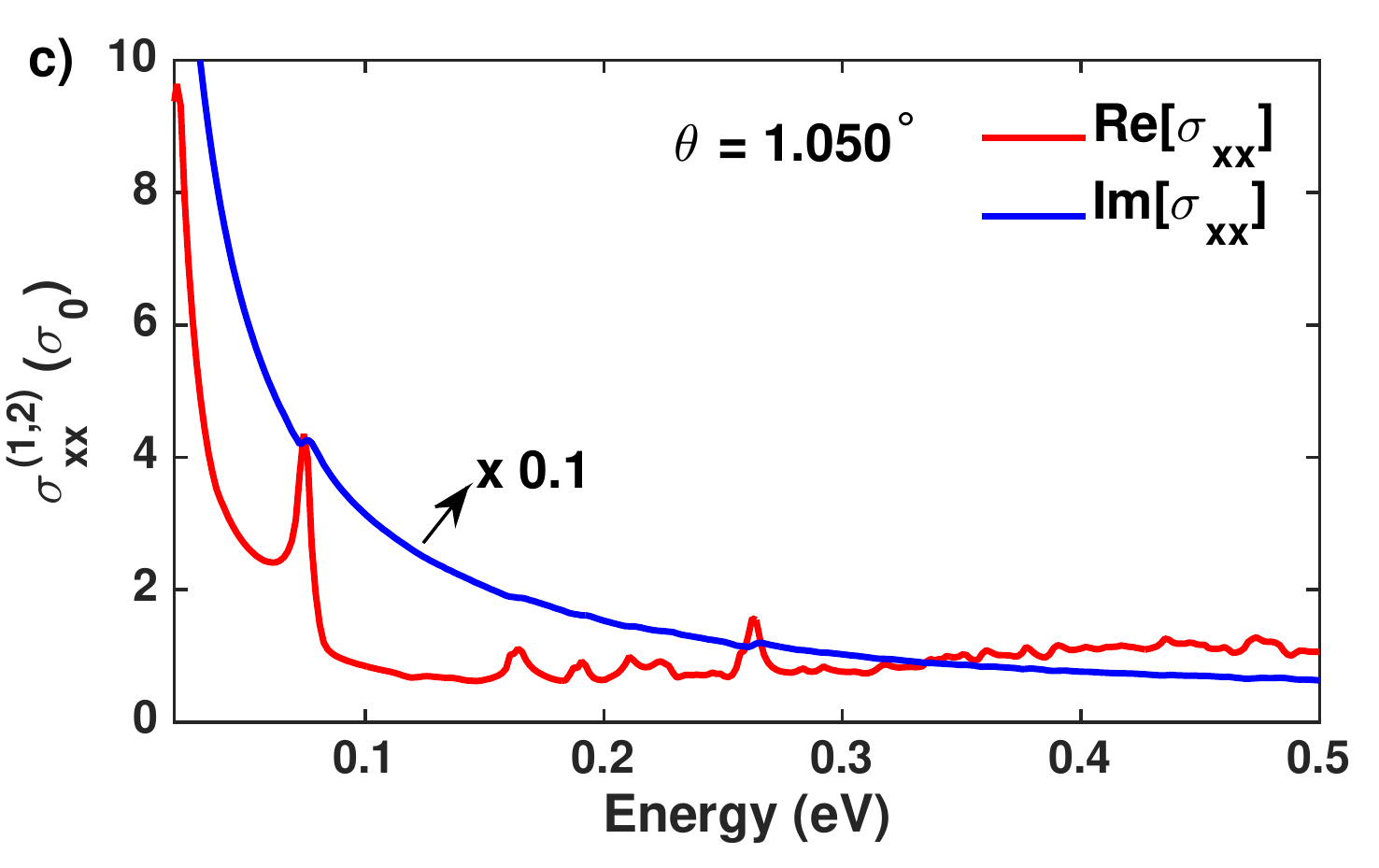}
\caption{\label{Fig4} These present the real and imaginary parts of the diagonal element $\sigma_{xx}^{(\ell)}$ of the tensor $\boldsymbol{\sigma}^{(\ell)}(\omega)$ ($\ell = 1, 2$)  of three TBG configurations with $\theta = 3.890^\circ, 1.890^\circ$ and $1.050^\circ$.}
\end{figure}

\subsection{AC conductivity and optical Hall drag conductivity}
In Fig. \ref{Fig3} we present the calculated results for the total optical conductivity of three TBG configurations. The calculation was performed on the use of the Kubo formula for the conductivity tensor wherein the velocity operator $\hat{v}_\alpha$ is determined from the Hamiltonian (\ref{Eq15}) using Eq. (\ref{Eq6}). In principle, the time reversal symmetry as well as the Onsager reciprocal relations rules out the off-diagonal elements of the conductivity tensor. We numerically checked this point and observed that the data of $\sigma_{xy}(\omega)$ behaves as a noise picture with the amplitude of $10^{-3}$, correctly verified the general principle. Additionally, all the symmetry properties of the conductivity tensor according to Eq. (\ref{Eq5}) are verified. It thus emphasizes the correctness of the numerically obtained data. Fig. \ref{Fig3} shows the longitudinal conductivity curves $\sigma_{xx}(\omega)$ with significant peaks, which are assigned to the dominant contribution of interband transition processes. This analysis is supported by associating it with the electronic band structures of the TBG configurations shown in Fig. \ref{Fig1}. For example, in Fig. \ref{Fig1}(b), at the energy position $E = 0.5681$ eV there appears an optical absorption peak that can be assigned to the transition from the highest valence band to the lowest conduction band at the value of $\vect{k}$ around the $M$ points in the mini-Brillouin zone. When decreasing the twist angle value, the picture of the electronic structure becomes more complicated. As a consequence, the optical conductivity shows more optical absorption peaks. For the TBG configurations with not too small twist angles ($> 2^\circ$), there is a noticeable feature of the conductivity curve in the low energy range: the real part of the conductivity behaves independently of the photon energy $\hbar\omega$ and takes the value of $2\sigma_0 $ where $\sigma_0 = \pi e^2/2h$. This behaviour is similar to that of monolayer graphene, which is pointed out to be a consequence of the linear dispersion law of electron states. In the low energy range, the energy band structure of the TBG systems qualitatively resembles that of monolayer graphene,\cite{Nicol_2008} but with a smaller Fermi velocity as analyzed previously. The value of  $\sigma_{xx}(\omega)$ is, however, equal to $2\sigma_0$.\cite{Do_2018,Do_2020,Moon-2013} For the case of magic angle $\theta = 1.050^\circ$, our calculation shows clearly the dominant contribution of the optical transitions from the quasi-flat bands around the Fermi level to the lowest conduction band around the $\Gamma$ point, see Fig. \ref{Fig3}(c) together with Figs. \ref{Fig1}(c,f).

\begin{figure*}\centering
\includegraphics[clip=true,trim=1.6cm 11cm 5cm 7cm,width=0.45\textwidth]{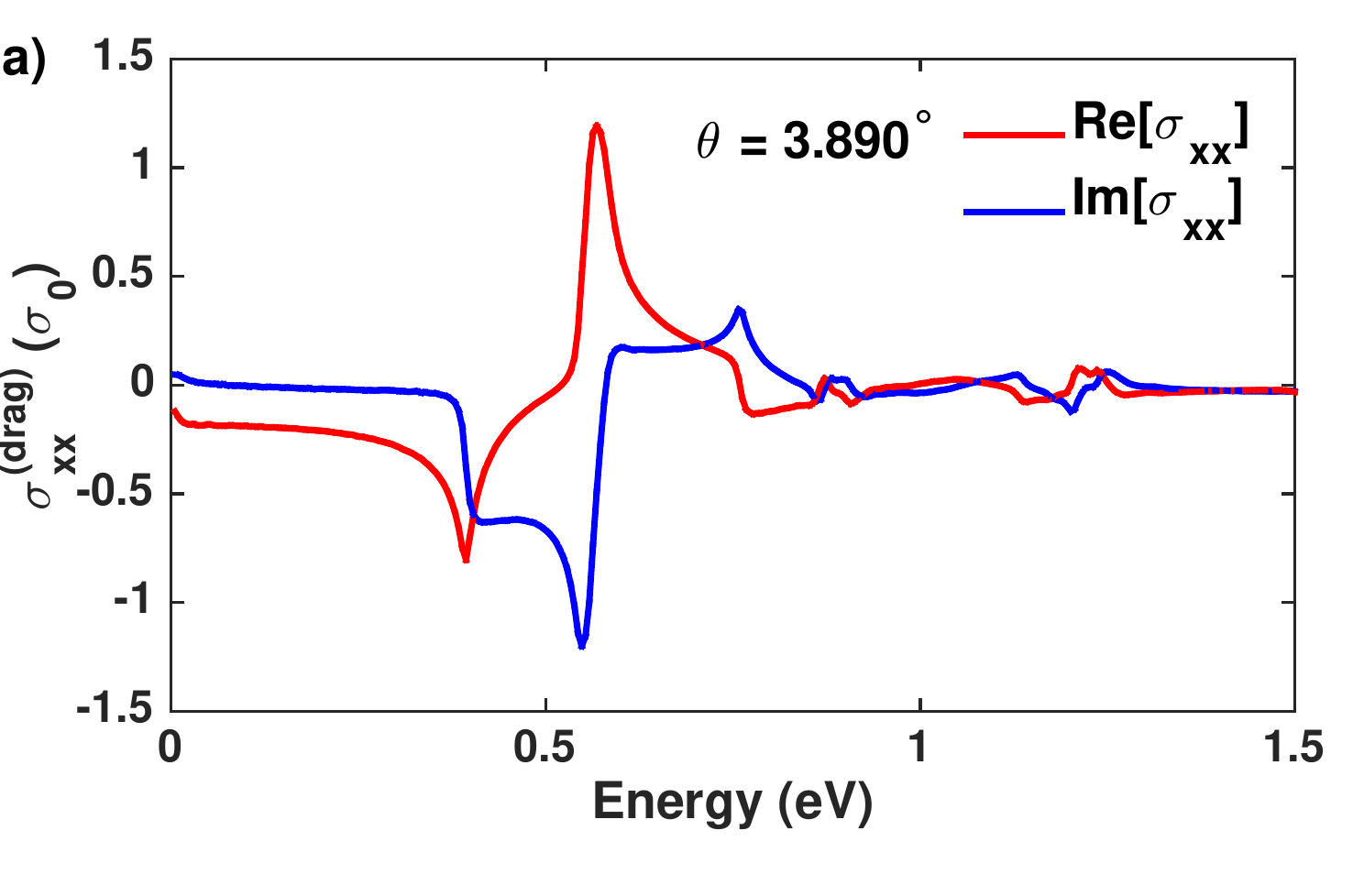}
\includegraphics[clip=true,trim=1.3cm 11cm 5.3cm 7cm,width=0.45\textwidth]{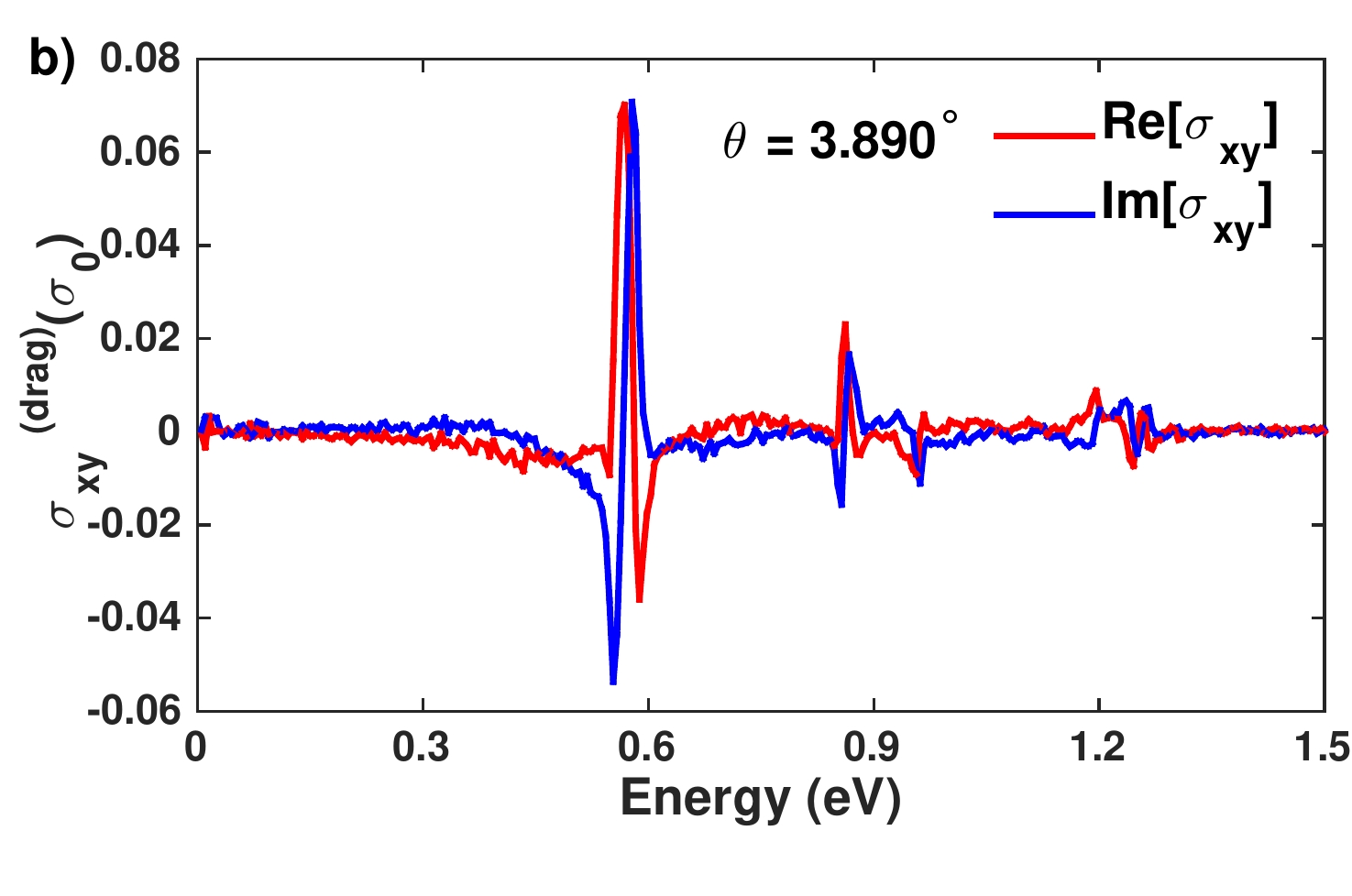}\\
\includegraphics[clip=true,trim=1.6cm 11cm 5cm 7cm,width=0.45\textwidth]{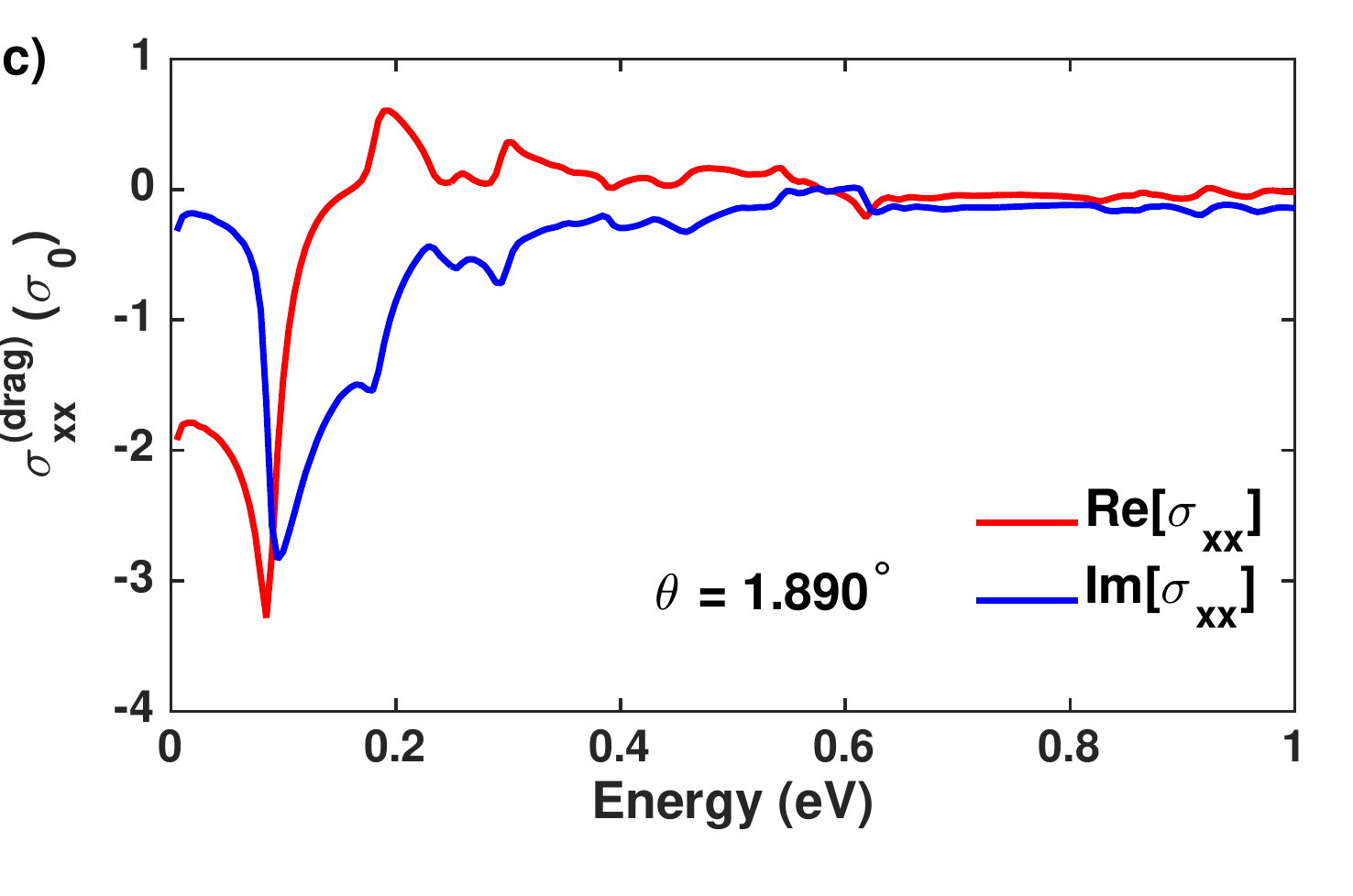}
\includegraphics[clip=true,trim=1.3cm 11cm 5.3cm 7cm,width=0.45\textwidth]{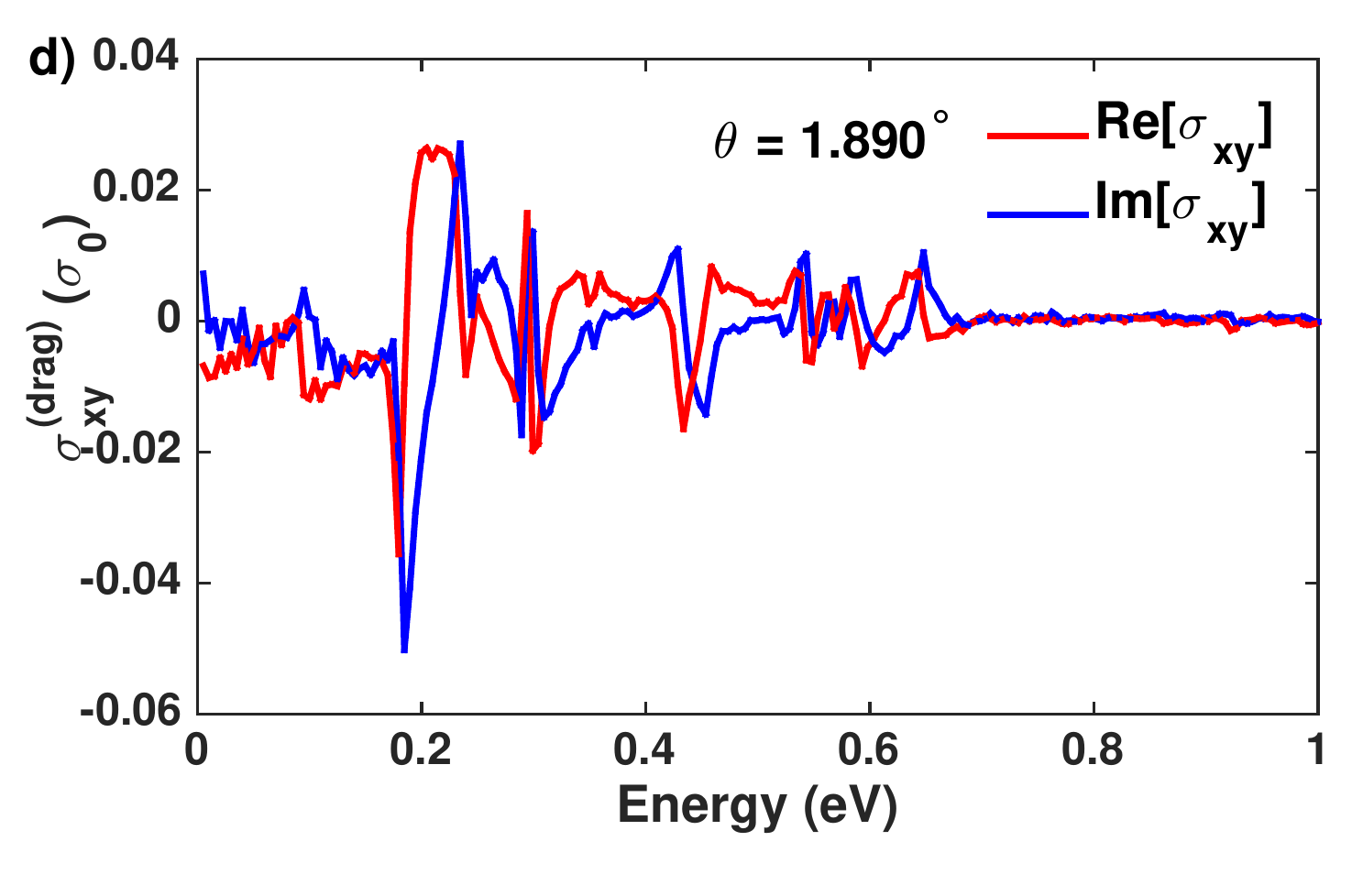}\\
\includegraphics[clip=true,trim=1.6cm 11cm 5cm 7cm,width=0.45\textwidth]{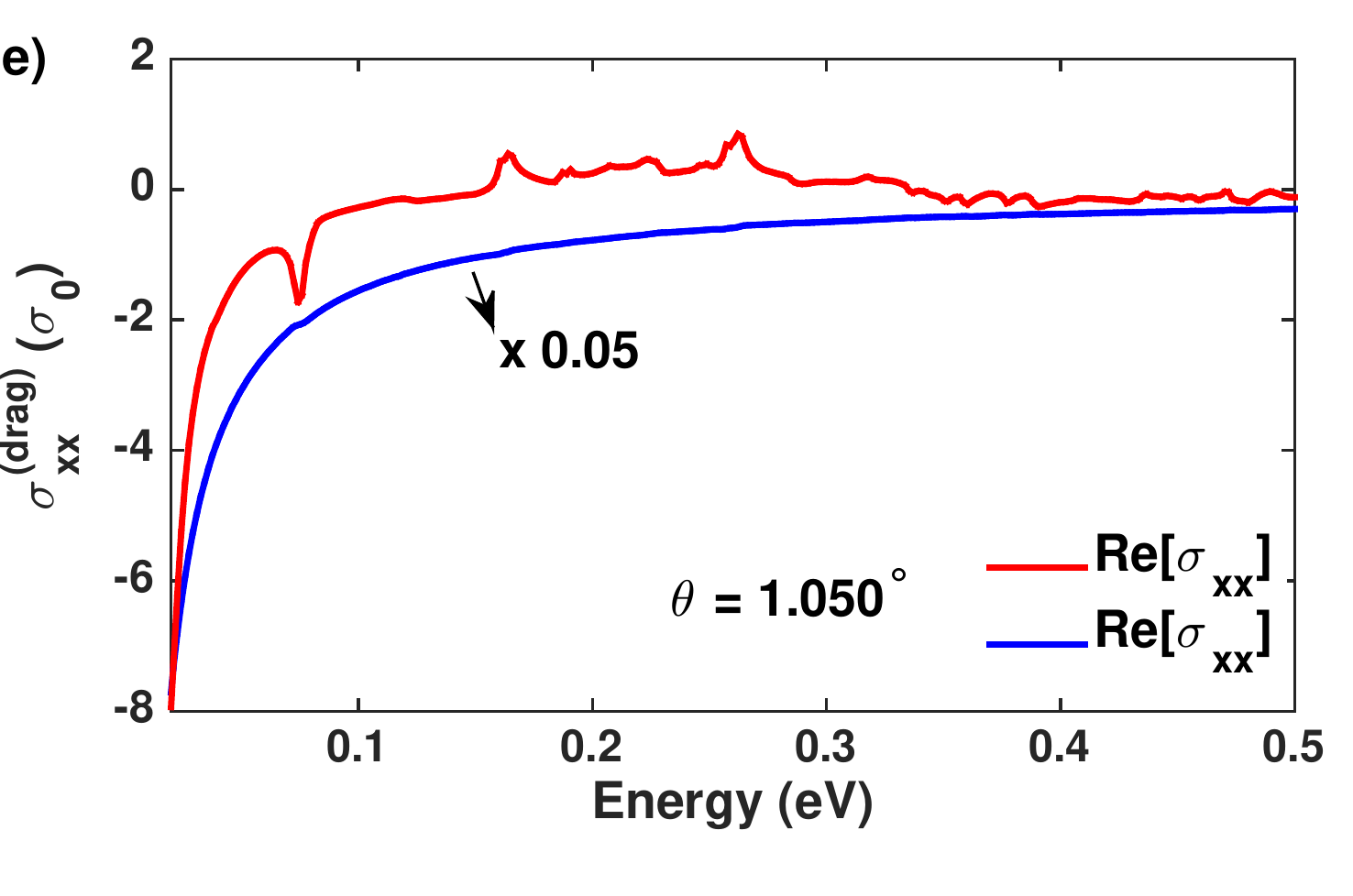}
\includegraphics[clip=true,trim=1.3cm 11cm 5.3cm 7cm,width=0.45\textwidth]{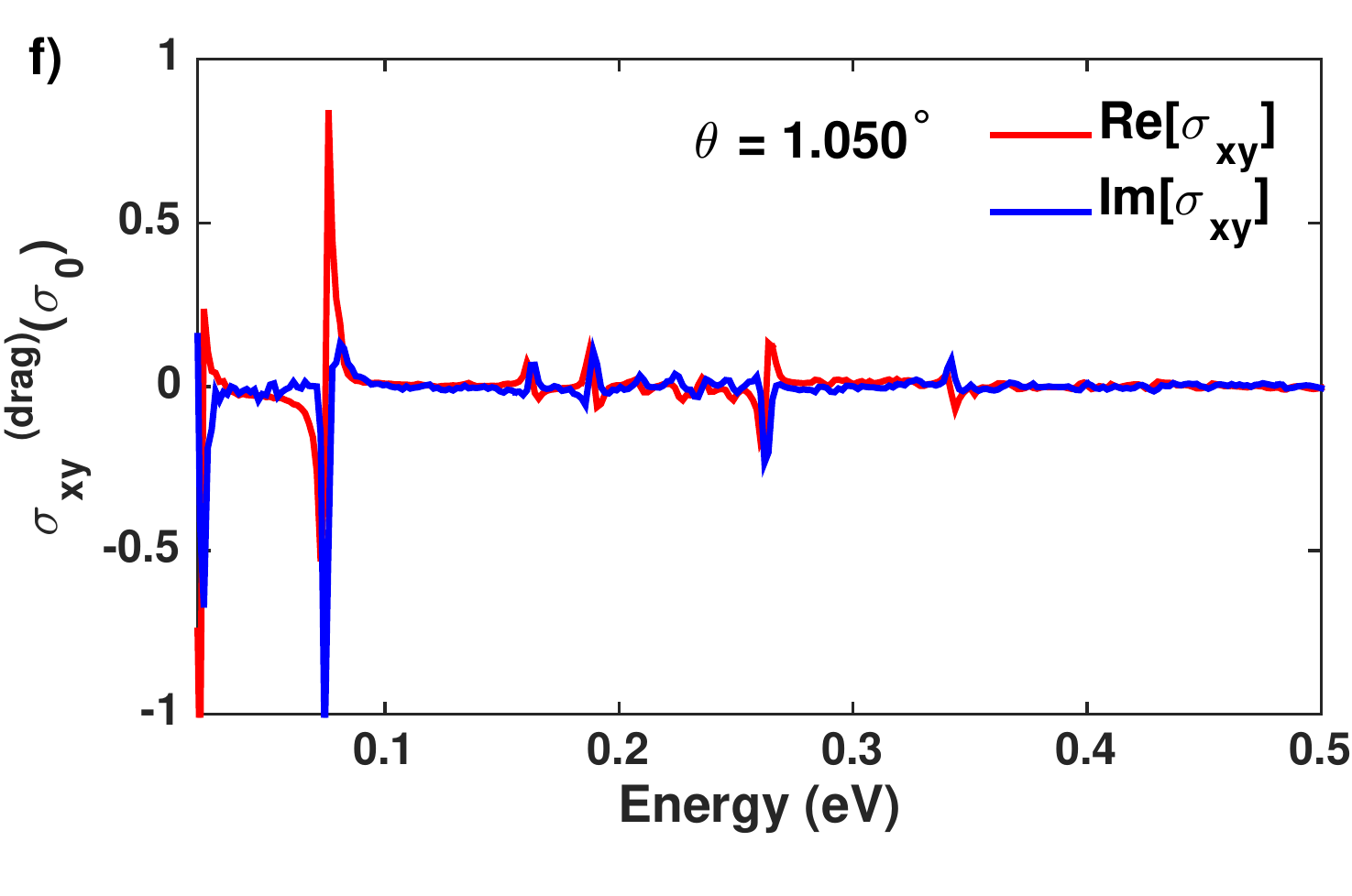}
\caption{\label{Fig5} These present the real and imaginary parts of the drag term $\boldsymbol{\sigma}^{(drag)}(\omega)$ of the conductivity tensor of three TBG configurations with $\theta = 3.890^\circ, 1.890^\circ$ and $1.050^\circ$: (a,c,e) for the longitudinal (diagonal) element $\sigma_{xx}^{(drag)}(\omega)$, and (b,d,f) for the transverse (off-diagonal) element $\sigma_{xy}^{(drag)}(\omega)$.}
\end{figure*}

In Fig. \ref{Fig4} and Fig. \ref{Fig5}, we present the calculation results for the elements of the blocks $\boldsymbol{\sigma}^{(\ell)}(\omega)$ and $\boldsymbol{\sigma}^{(drag)}(\omega)$ of the conductivity tensor according to the analysis given in Sec. II.A. Our numerical calculation obviously shows $\boldsymbol{\sigma}^{(1)}(\omega) = \boldsymbol{\sigma}^{(2)}(\omega) = \sigma_{xx}^{(\ell)}(\omega)\vect{\tau}_0$. This result verifies the equivalent role of two graphene layers in the TBG lattice. Figs. \ref{Fig4}(a,b,c) shows the real part (red) and the imaginary part (blue) of the element $\sigma_{xx}^{(\ell)}(\omega)$ where $\ell = 1, 2$. One noticeable point is that when decreasing the value of the twist angle, the magnitude of $\sigma_{xx}^{(\ell)}$ increases; especially, the imaginary part becomes dominant and varying via the law of $1/\hbar\omega$, see Fig. \ref{Fig4}(c). Next, Fig. \ref{Fig5} shows the real and imaginary parts of the off-diagonal element $\boldsymbol{\sigma}^{(drag)}(\omega)$ of the drag part of the conductivity tensor. For the TBG configurations of large twist angle, there is a significant difference between the curves $\sigma_{xx}^{(\ell)}(\omega)$ and $\sigma_{xx}^{(drag)}(\omega)$, see Fig. \ref{Eq4}(a) and Fig. \ref{Fig5}(a). However, for the configurations with tiny twist angles ($< 2^\circ$), these curves are mostly the same but they are opposite in sign. Accordingly, though there is a large variation of the magnitude of $\sigma_{xx}^{(\ell)}(\omega)$ and $\sigma_{xx}^{(drag)}(\omega)$, they compensate each other in the summation to result in the total value $\sigma_{xx}(\omega) = 2[\sigma_{xx}^{(\ell)}(\omega)+\sigma_{xx}^{(drag)}(\omega)]$ as shown in Fig. \ref{Fig2}. For the behaviour of $\sigma_{xy}^{(drag)}(\omega)$, it mainly shows the zero-close value in almost energy range except for some sharp peaks and dips at the energy positions of strong optical absorption.

\begin{figure}\centering
\includegraphics[clip=true,trim=1.5cm 6.5cm 2.5cm 7cm,width=\columnwidth]{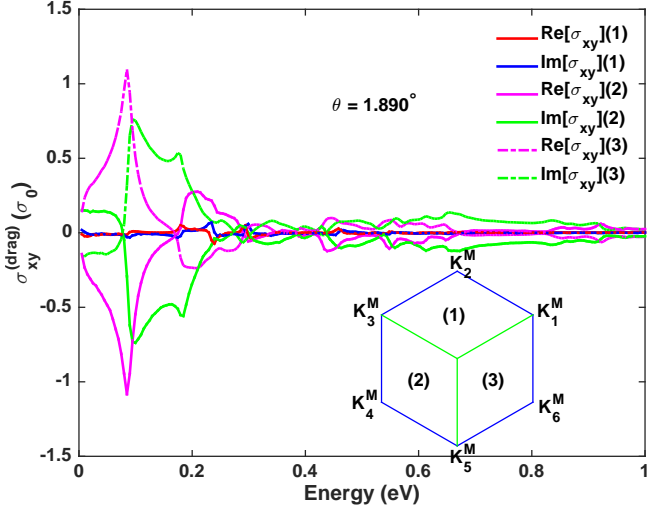}
\caption{\label{Fig6} This presents the contribution of Bloch states to $\sigma_{xy}^{(drag)}(\omega)$: The red and blue curves are from the Bloch states with $\vect{k}$ in the first one-third Brillouin zone; the solid/dashed purpe and green curves are from the Bloch states with $\vect{k}$ in the second/third one-third Brillouin zone.}
\end{figure}

The appearance of the transverse conductivity $\sigma_{xy}^{(drag)}(\omega)$ is strikingly important. This is because it governs typical optical properties of the TBG systems. This conductivity does not result from only the correlation of current densities distributed on two graphene layers, but also the chirality of the atomic lattices. The consequence of the latter is that the TBG atomic lattices have no mirror symmetries, totally different from the hexagonal lattice of graphene. The studies made by Morell and by Do show that the hybridized states formed in the bilayer systems always support a finite value of the correlation of two transverse components of electron motion. In other words, a motion of electron in one graphene layer along the $Ox$ direction always induces another motion in the second graphene layer along the transverse direction $Oy$ due to the interlayer coupling. However, the contribution of such correlations to the total value of the conductivity must cancel each other completely if the system has a mirror symmetry. Since the TBG lattices are chiral, such cancellation should be incomplete and thus results in the finite non-zero conductivity $\sigma_{xy}^{(drag)}(\omega)\neq 0$. In order to emphasize this point, we divided the hexagonal mini-Brillouin zone of the TBG lattice into three equal parts as shown in the inset of Fig. \ref{Fig6}. The division is based on the assumption that there is a mirror symmetry via the plane $M_{yz}$. Geometrically, the domain (1) would contain the symmetry $M_{yz}$ meanwhile the domains (2) and (3) would be the mirror images of each other. Using the Kubo formula we calculated the value of $\sigma_{xy}^{(drag)}$ by taking the summation over all the $\vect{k}$ points in these three domains separately. Obtained results are shown in Fig. \ref{Fig6}. We clearly see the results obtained in the domains (2) and (3) are very similar to each other, but opposite in sign, and much larger than that obtained in the domain (1). Obviously, when taking into account the summation of the three results, a strong, but not complete, cancellation of the data occurs and results in the final value of $\sigma_{xy}^{(drag)}(\omega)$ as shown in Figs. \ref{Fig5}(b,d,f). For the two special AA- and AB-stacked configurations, which processes the mirror symmetries, we verified that the cancellation takes place completely, and thus $\sigma_{xy}^{(drag)}(\omega)=0$. We call this result the chiral correlation of drag current densities distributed in the two graphene layers. The essential role of the chirality of the TBG atomic lattices is therefore demonstrated. However, we should emphasize that its essence lies in governing symmetrical properties of electronic states. This can be seen as the sufficient condition for   $\sigma_{xy}^{(drag)}(\omega)\neq 0$. Meanwhile, the necessary condition must be the electronic interlayer coupling, which results in the formation of electronic states that always support the drag correlation. The Kubo formula allows to calculate each elements of the conductivity tensor independently. Technically, this can be seen as the result of decomposing the total velocity operator into appropriate terms, similar to Morell's strategy. However, it should be kept in mind that the calculation of the velocity-velocity correlation is actually realized on the basis of the electronic states of the TBG systems, not of Bloch states of individual graphene.

\begin{figure}\centering
\includegraphics[clip=true,trim=1.5cm 6.5cm 2.5cm 7cm,width=\columnwidth]{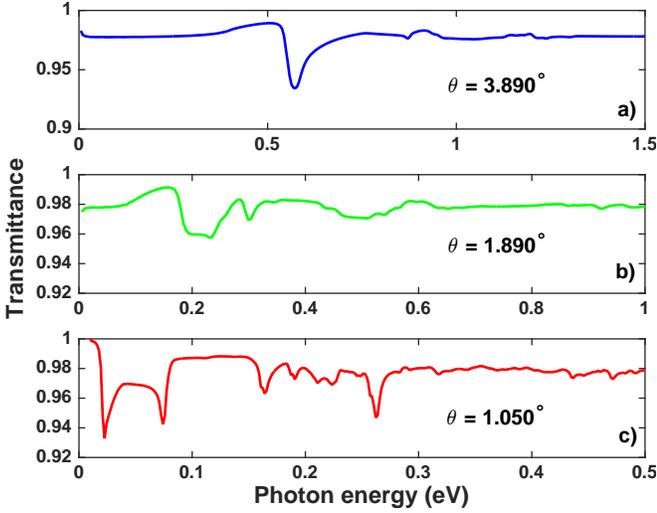}
\caption{\label{Fig7} This presents the transmission spectra of three TBG configurations: $\theta = 3.890^\circ$ (a), $\theta = 1.890^\circ$ (b), and $\theta = 1.050^\circ$ (c).}
\end{figure}

\begin{figure}\centering
\includegraphics[clip=true,trim=1.3cm 6.5cm 2.5cm 7cm,width=\columnwidth]{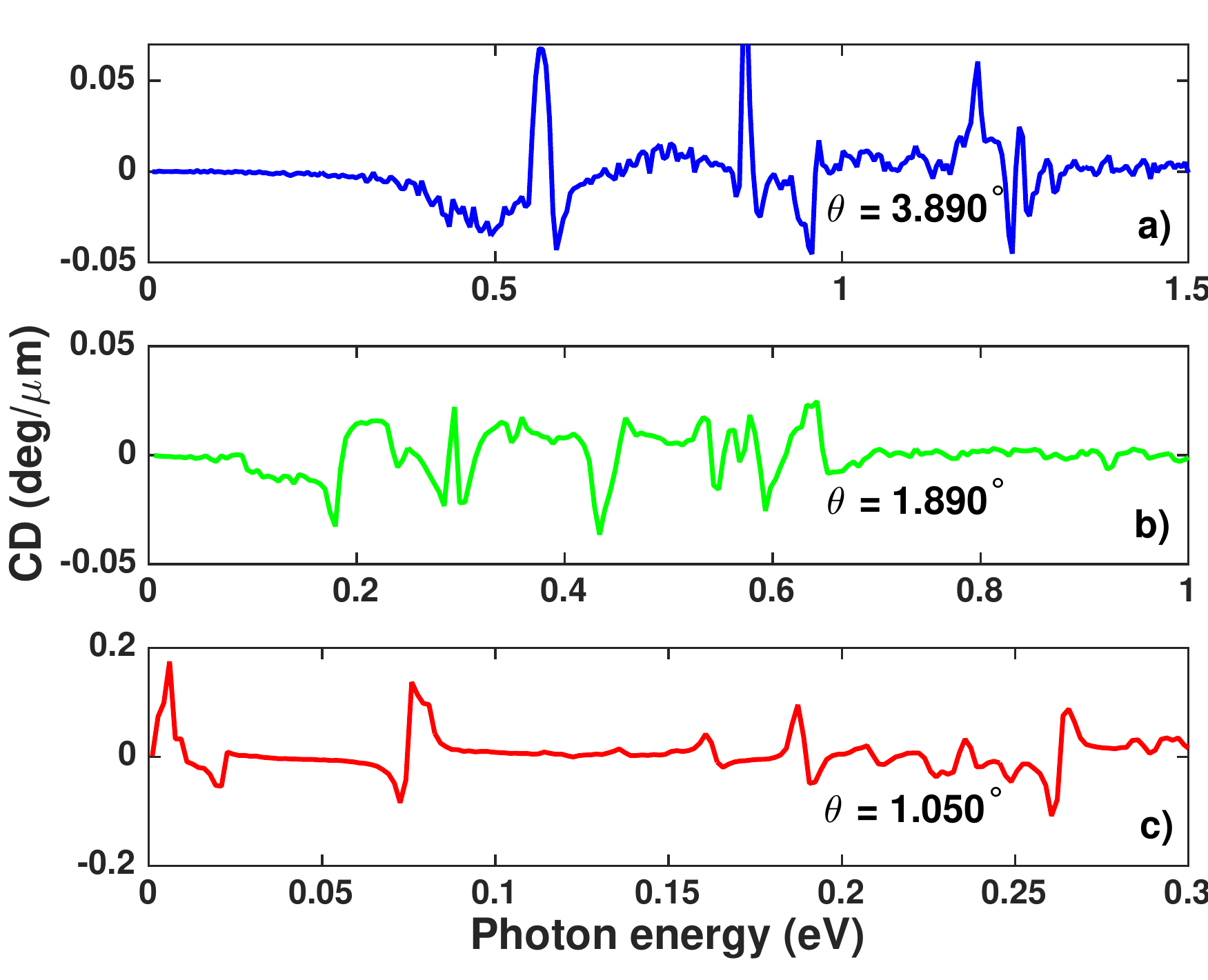}
\caption{\label{Fig8} This presents the circular dichroism (CD) spectra of three TBG configurations: $\theta = 3.890^\circ$ (a), $\theta = 1.890^\circ$ (b), and $\theta = 1.050^\circ$ (c).}
\end{figure}

\subsection{Optical chiral responses}
In order to investigate the optical response of the TBG systems, in this section, we first establish a theory for the problem of electromagnetic wave propagation through a generic TBG sheet whose thickness $d_{GG}$ is not ignored. Different from an ordinary dielectric slap with two independent surfaces, the electric current densities in two graphene layers are mutually related in the meaning that: a current density in one graphene layer is induced by the value of electric field not only in that layer, but also in the other layer. This relation is given by Eq. (\ref{Eq4}). We used this equation to solve the problem defined here.

Specifically, we set up the system like the one in Ref.~\onlinecite{ Szechenyi_2016} in which two graphene layers separates the whole space into three parts, which are characterized by the parameters $(\epsilon_i,\mu_i)$, where $i = 1, 2, 3$ is the index labeling the mediums. Our aim is to establish the transfer matrix of the whose system in order to build the expression for the transmission and reflection matrices for the system of bilayer graphene. Since the whole space is only inhomogeneous in the direction perpendicular to the TBG surface, called the $z$ direction, the Maxwell equations support a plane wave solution that is given by the expression:
\begin{equation}\label{Eq25}
    \vect{F}_i(x,y,z) = \left[\vect{F}_i^+e^{ik_{iz}}+\vect{F}_i^{-}e^{-ik_{iz}z}\right]e^{ik_{ix}x}e^{ik_{iy}y},
\end{equation}
where $\vect{F}$ means $\vect{E}$ (the electric field), $\vect{D}$ (the electric induction), $\vect{H}$ (the magnetic field) and $\vect{B}$ (the magnetic induction). The vector coefficients $\vect{F}_i^\pm$ depend on the wave vector $\vect{k}_i$ in general. Conditions required for the continuity of the vector fields at the interface between two mediums are given by:
\begin{subequations}
\begin{align}
    &\vect{n}\times[\vect{E}_2(1)-\vect{E}_1(1)] = 0,\label{Eq26a}\\
    &\vect{n}\times[\vect{H}_2(1)-\vect{H}_1(1)] = \vect{J}(1),\label{Eq26b}
\end{align}
\end{subequations}
for the surface of the first graphene layer, and
\begin{subequations}
\begin{align}
    &\vect{n}\times[\vect{E}_3(2)-\vect{E}_2(2)] = 0,\label{Eq27a}\\
    &\vect{n}\times[\vect{H}_3(2)-\vect{H}_2(2)] = \vect{J}(2).\label{Eq27b}
\end{align}
\end{subequations}
for the surface of the second graphene layer. In Eqs. (\ref{Eq26b}) and (\ref{Eq27b}), the distribution of current densities on each graphene layer is linearly related to the electric field by the conductivity tensor given via Eq. (\ref{Eq5}). To the completeness of the setup, we rewrite this equation in the explicit form:
\begin{subequations}
\begin{align}
    \vect{J}(1) &= \vect{\sigma}^{(1)}\vect{E}_1(1)+\vect{\sigma}^{(drag)}\vect{E}_2(2),\label{Eq28a}\\
    \vect{J}(2) &= \vect{\sigma}^{(drag)\dagger}\vect{E}_1(1)+\vect{\sigma}^{(2)}\vect{E}_2(2)\nonumber\\
    &=\vect{\sigma}^{(drag)\dagger}\vect{E}_2(1)+\vect{\sigma}^{(2)}\vect{E}_3(2).\label{Eq28b}
\end{align}
\end{subequations}

The equations of continuity conditions allow us to establish the relation between the vector fields $(\vect{F}_i^+,\vect{F}_i^-)$ in two mediums at their interface. Particularly, we first notice that  the magnetic field $\vect{H}_i$ is related to the electric field $\vect{E}_i$ via the Maxwell equation $\text{curl}(\vect{E}) = -\partial_t\vect{B}$ wherein $\vect{B}=\mu\vect{H}$. Using the plane wave solution (\ref{Eq25}) we obtain:
\begin{equation}\label{Eq29}
    \vect{H}_i^\pm = \pm\sqrt{\frac{\epsilon_i}{\mu_i}}\frac{\vect{k}_i\times\vect{E}_i^\pm}{\|\vect{k}_i\|},
\end{equation}
where $\epsilon_i$ and $\mu_i$ are the absolute permittivity and the absolute permeability of medium $i$. In order to obtain the above equation, the dispersion relation $\omega = c\|\vect{k}_i\|/n_i$ is already used, where $c$ is the light speed in vacuum and $n_i$ is the refractive index of medium $i$. Notice that the value of  $\sqrt{\epsilon_i/\mu_i}$ can be expressed through $n_i$, i.e., $\sqrt{\epsilon_i/\mu_i} \approx n_i\sigma_0/2\alpha$ where   $\sigma_0=e^2/h$ is the unit of quantum conductivity and $\alpha = e^2/(4\pi\epsilon_0\hbar c)$ is the fine-structure constant. We solve the problem for the transmission of light along the normal direction to the TBG layer, so $\vect{k}_i = k_i\vect{n}$. It results in the transfer matrix $\mathbf{M}_{31}$ (of size $4\times 4$) relating the components of the electric fields in the first and third mediums
\begin{align}\label{Eq30}
    \left(\begin{array}{c}
         \vect{E}_3^+(2)  \\
         \vect{E}_3^-(2)
    \end{array}\right)
    &= \mathbf{M}_{31} \left(\begin{array}{c}
         \vect{E}_1^+(1)  \\
         \vect{E}_1^-(1)
    \end{array}\right).
\end{align}
Here $\mathbf{M}_{31}=\mathbf{M}_{32}\mathbf{M}^f_2\mathbf{M}_{21}$ wherein:
\begin{widetext}
\begin{subequations}
\begin{align}
    \mathbf{M}_{32} &= \left(\begin{array}{cc}
        \vect{1} & \vect{1} \\
        -\sqrt{\frac{\epsilon_3}{\mu_3}}\vect{1}-\vect{\sigma}^{(2)} &\sqrt{\frac{\epsilon_3}{\mu_3}}\vect{1}-\vect{\sigma}^{(2)} 
    \end{array}\right)^{-1}\left(\begin{array}{cc}
        \vect{1} & \vect{1} \\
        -\sqrt{\frac{\epsilon_2}{\mu_2}}\vect{1}+\vect{\sigma}^{(drag)\dagger}e^{-ik_{2z}d} & \sqrt{\frac{\epsilon_2}{\mu_2}}\vect{1}+\vect{\sigma}^{(drag)\dagger}e^{ik_{2z}d} 
    \end{array}\right), \label{Eq31a}\\
        \mathbf{M}_{21} &= \left(\begin{array}{cc}
    \vect{1}&\vect{1}\\
    -\sqrt{\frac{\epsilon_2}{\mu_2}}\vect{1}-\vect{\sigma}^{(drag)}e^{ik_{2z}d}&\sqrt{\frac{\epsilon_2}{\mu_2}}\vect{1}-\vect{\sigma}^{(drag)}e^{-ik_{2z}d}\end{array}\right)^{-1}\left(\begin{array}{cc}
      \vect{1}   & \vect{1} \\
    -\sqrt{\frac{\epsilon_1}{\mu_1}}\vect{1}+\vect{\sigma}^{(1)}     & \sqrt{\frac{\epsilon_1}{\mu_1}}\vect{1}+\vect{\sigma}^{(1)}
    \end{array}\right), \label{Eq31b}\\
    \mathbf{M}_2^f &=\left(\begin{array}{cc}
    e^{ik_{2z}d}\vect{1} & 0 \\
        0 & e^{-ik_{2z}d}\vect{1} 
    \end{array}\right), \label{Eq31c}
\end{align}
\end{subequations}
\end{widetext}
where $\vect{1}$ is the $2\times 2$ identity matrix.
Let $\mathbf{r}$ and $\mathbf{t}$ be the $2\times 2$ matrices relating the components of the electric field vector of the incident light to those of the reflected and transmitted light, i.e., $\vect{E}_1^-(1) = \mathbf{r}\cdot\vect{E}_1^+$ and $\vect{E}_3^+=\mathbf{t}\cdot\vect{E}_1^+$. These matrices are called the reflection and transmission matrices, respectively. From Eq. (\ref{Eq30}) we deduce:
\begin{subequations}
\begin{align}
    \mathbf{r} &= -[\mathbf{M}_{31}^{22}]^{-1}\mathbf{M}_{31}^{21}, \label{Eq32a}\\
    \mathbf{t} &= \mathbf{M}_{31}^{11}-\mathbf{M}_{31}^{12}[\mathbf{M}_{31}^{22}]^{-1}\mathbf{M}_{31}^{21}=[\mathbf{M}_{31}^{-1}]^{11}. \label{Eq32b}
\end{align}
\end{subequations}
From these matrices, the physical quantities as the reflectance ($R$), transmittance ($T$) are determined as the ratio of the energy fluxes. Concretely,
\begin{subequations}
\begin{align}
    R &= \left\vert\frac{(\mathbf{r}\cdot\vect{E}_1^+)^\dagger\cdot(\mathbf{r}\cdot\vect{E}_1^+)}{\vect{E}_1^{+\dagger}\cdot\vect{E}_1^+}\right\vert, \label{Eq33a}\\
    T &= \left\vert\frac{n_3}{n_1}\frac{(\mathbf{t}\cdot\vect{E}_1^{+})^\dagger\cdot(\mathbf{t}\cdot\vect{E}_1^+)}{\vect{E}_1^{+\dagger}\cdot\vect{E}_1^+}\right\vert. \label{Eq33b}
\end{align}
\end{subequations}
From the conservation of the energy flux, the absorptance is determined by $A = 1-(R+T)$. The quantities $R,T$ and $A$ depend on not only the energy but also the polarization states of light. Remember that the left/right-handed polarization states of light are characterized by the electric field vectors $\vect{E}_{L,R}^+ \propto (1,\pm i)^T/\sqrt{2}$. We calculated the spectrum of circular dichroism (CD), a quantity is defined to measure the dependence of the light absorption of the left-handed (L) and right-handed (R) polarization by the formula:
\begin{equation} \label{Eq34}
\mathrm{CD} = \frac{A_\mathrm{L}-A_\mathrm{R}}{A_\mathrm{L}+A_\mathrm{R}},
\end{equation}

On the basis of the theory established above and the data of the conductivity tensor already computed, we performed the calculation for the transfer matrix, then the reflection, transmission and absorption spectra of several TBG configurations. Obtained results are presented in Fig. \ref{Fig7} for the transmittance and in Fig. \ref{Fig8} for the CD. The reflectance is very small, so we ignore to show it here. From the transmission spectra, we see that the transmittance of all TBG configurations can get the value of 98\% and the absorptance of 2\% in average in a large energy range. The spectral curves clearly exhibit the peaks and dips consistent with the picture of the longitudinal conductivities. The transmission and absorption spectra for the left- and right-handed lights are different to be distinguished. However, the CD spectra is clearly manifested. From Fig. \ref{Fig8}, we see that for the TBG configuration with $\theta = 3.890^\circ$,  the CD curve exhibits essential features of experimental data reported by Kim et. al.\cite{Kim_2016}, for instance, with the peaks and dips of large width, except several sharp peaks at the energy position of the absorption peaks. In order to verify the role of the drag Hall conductivity $\sigma_{xy}^{(drag)}$, we did not include it into the calculation. As expected, it results in the zero CD for the total energy range, but mostly does not change the transmission spectrum. We therefore conclude that this drag Hall conductivity plays the decisive role in governing the chiral optical behaviors of the TBG systems.

\section{Conclusion}
Engineering two-dimensional quantum materials to electronic and opto-electronic applications is the research direction that has been extensively developing. In this work, we have presented a study of the mechanism of optical activity of twisted bilayer graphene, a typical two-dimensional van der Waals material. The important feature of this material system is that its atomic lattice processes only rotational symmetries, but not the mirror symmetry due to the chiral structure. We used an effective continuum model to show the hybridization of electron states in each graphene layer to form unique electron states in the bilayer system. Such hybridized states support the correlation of transverse motions in two graphene layers. We established a solution to the problem of electromagnetic wave propagation through a TBG sheet wherein each graphene layer is considered as the conducting interface separating two dielectric mediums. The two interfaces are not independent, but they conduct the mutually influenced current densities. By this, we considered explicitly the spatial dispersion and presented a theory analyzing the conductivity tensor into two parts, the local and the drag ones. We calculated all the elements of these parts using the Kubo formula. We then formulated and numerically calculated the transfer matrix and demonstrated that the transverse drag Hall conductivity $\sigma_{xy}^{(drag)}$ plays the decisive role in defining the chiral optical response of the TBG systems. We showed the existence of $\sigma_{xy}^{(drag)}$ due to the chiral structure of the atomic lattice. This results in the hybridized states of lacking the mirror symmetry. Such states not only support the correlation of transverse motions in two graphene layers, but also leads to the incomplete cancellation of such drag correlation to form this drag Hall optical conductivity. Another important result in this work is what we calculated the DC conductivity of the TBG systems and showed the existence of a quantum conductivity value $\propto e^2/h$ at the intrinsic Fermi energy. This quantum conductivity value is governed by the hybridized states in the bilayer systems. Our study therefore emphasises that in order to correctly analyze the optical response and transport properties, the finite thickness of the twisted bilayer graphene must be taken into account because of the importance of spatial dispersion effects. The theoretical approach we developed here is suitable to utilize for other van der Waals material systems of multiple layers.

\section*{Acknowledgements}
The authors are grateful to Shingo Takeuchi for reading carefully the manuscript before it was submitted. One of the authors, S.T.H., acknowledges Hanoi University of Civil Engineering (HUCE) for financially funding his work under the grant numbered 28-2021/KHXD-TD.

\bibliographystyle{apsrev4-1}
\bibliography{bibliography}

\end{document}